\documentclass[11pt]{article}%
\usepackage{amsfonts}
\usepackage{amssymb}
\usepackage{amsmath}
\usepackage{amscd}
\usepackage{latexsym}
\usepackage[onehalfspacing]{setspace}
\usepackage{theorem}
\usepackage{natbib}
\usepackage{hyperref}
\usepackage{graphicx}
\usepackage{caption}
\usepackage{subfig}
\hypersetup{pdfborder = {0 0 0},colorlinks=true,linkcolor=blue,urlcolor=blue,citecolor=blue}

\pdfoutput=1

\providecommand{\U}[1]{\protect\rule{.1in}{.1in}}
\newtheorem{theorem}{Theorem}

\newtheorem{Assumption}{Assumption}
\newcommand{\I}{{\rm 1\hspace*{-0.4ex}\rule{0.1ex}{1.52ex}\hspace*{0.2ex}}}
\hypersetup{pdfborder = {0 0 0},colorlinks=true,linkcolor=blue,citecolor=blue}
\renewcommand{\cite}{\citet*}

\begin{document}

\title{Simple Local Polynomial Density Estimators\thanks{A preliminary version of
this paper circulated under the title \textquotedblleft Simple Local
Regression Distribution Estimators with an Application to Manipulation
Testing\textquotedblright. We thank Sebastian Calonico, Toru Kitagawa, Zhuan
Pei, Rocio Titiunik, Gonzalo Vazquez-Bare, the co-editor, Hongyu Zhao, an
associate editor, and two reviewers for useful comments that improved our work
and software implementations. Cattaneo gratefully acknowledges financial
support from the National Science Foundation (SES 1357561 and SES 1459931).
Jansson gratefully acknowledges financial support from the National Science
Foundation (SES 1459967) and the research support of CREATES (funded by the
Danish National Research Foundation under grant no. DNRF78).}\bigskip}
\author{Matias D. Cattaneo\thanks{Department of Operations Research and Financial
Engineering, Princeton University.}
\and Michael Jansson\thanks{Department of Economics, UC Berkeley and \emph{CREATES}%
.}
\and Xinwei Ma\thanks{Department of Economics, UC San Diego.}}
\maketitle

\begin{abstract}
This paper introduces an intuitive and easy-to-implement nonparametric density
estimator based on local polynomial techniques. The estimator is fully
boundary adaptive and automatic, but does not require pre-binning or any other
transformation of the data. We study the main asymptotic properties of the
estimator, and use these results to provide principled estimation, inference,
and bandwidth selection methods. As a substantive application of our results,
we develop a novel discontinuity in density testing procedure, an important
problem in regression discontinuity designs and other program evaluation
settings. An illustrative empirical application is given. Two companion
\texttt{Stata} and \texttt{R} software packages are provided.\bigskip

\end{abstract}

\bigskip%

%TCIMACRO{\TeXButton{setup}{\thispagestyle{empty}
%\setcounter{page}{0}
%\newpage\doublespacing}}%
%BeginExpansion
\thispagestyle{empty}
\setcounter{page}{0}
\newpage\doublespacing
%EndExpansion
%

%TCIMACRO{\TeXButton{SET PATH HERE}{\newcommand{\pathpaper}%
%{PUT-LAST-WHAT-YOU-WANT}
%\renewcommand{\pathpaper}%
%{C:/Documents/Research/Cattaneo-Jansson-Ma--LocPolDensity}
%\renewcommand{\pathpaper}{Z:/research/Cattaneo-Jansson-Ma_2019_JASA}}}%
%BeginExpansion
\newcommand{\pathpaper}{PUT-LAST-WHAT-YOU-WANT}
\renewcommand{\pathpaper}%
{C:/Documents/Research/Cattaneo-Jansson-Ma--LocPolDensity}
\renewcommand{\pathpaper}{Z:/research/Cattaneo-Jansson-Ma_2019_JASA}%
%EndExpansion

\section{Introduction\label{section:intro}}

Flexible (nonparametric) estimation of a probability density function features
prominently in empirical work in statistics, economics, and many other
disciplines. Sometimes the density function is the main object of interest,
while in other cases it is a useful ingredient in forming two-step
nonparametric or semiparametric procedures. In program evaluation and causal
inference settings, for example, nonparametric density estimators are used for
manipulation testing, distributional treatment effect and counterfactual
analysis, instrumental variables treatment effect specification and
heterogeneity analysis, and common support/overlap testing. See
\cite{Imbens-Rubin_2015_Book} and \cite{Abadie-Cattaneo_2018_ARE} for reviews
and further references.

A common problem faced when implementing density estimators in empirical work
is the presence of evaluation points that lie on the boundary of the support
of the variable of interest: whenever the density estimator is constructed at
or near boundary points, which may or may not be known by the researcher, the
finite- and large-sample properties of the estimator are affected. Standard
kernel density estimators are invalid at or near boundary points, while other
methods may remain valid but usually require choosing additional tuning
parameters, transforming the data, a priori knowledge of the boundary point
location, or some other boundary-related specific information or modification.
Furthermore, it is usually the case that one type of density estimator is used
for evaluation points at or near the boundary, while a different type is used
for interior points.

We introduce a novel nonparametric estimator of a density function constructed
using local polynomial techniques \citep{Fan-Gijbels_1996_Book}. The estimator
is intuitive, easy to implement, does not require pre-binning of the data, and
enjoys all the desirable features associated with local polynomial regression
estimation. In particular, the estimator automatically adapts to the
boundaries of the support of the density without requiring specific data
modification or additional tuning parameter choices, a feature that is
unavailable for most other density estimators in the literature: see
\cite{Karunamuni-Alberts_2005_SM} for a review on this topic. The most closely
related approaches currently available in the literature are the local
polynomial density estimators of \cite{Cheng-Fan-Marron_1997_AoS} and
\cite{Zhang-Karunamuni_1998_JSPI}, which require knowledge of the boundary
location and pre-binning of the data (or, more generally, pre-estimation of
the density near the boundary), and hence introduce additional tuning
parameters that need to be chosen.

The heuristic idea underlying our estimator, and differentiating it from other
existing ones, is simple to explain: whereas other nonparametric density
estimators are constructed by smoothing out a histogram-type estimator of the
density, our estimator is constructed by smoothing out the empirical
distribution function using local polynomial techniques. Accordingly, our
density estimator is constructed using a preliminary tuning-parameter-free and
$\sqrt{n}$-consistent distribution function estimator (where $n$ denotes the
sample size), implying in particular that the only tuning parameter required
by our approach is the bandwidth associated with the local polynomial fit at
each evaluation point. For the resulting density estimator, we provide (i)
asymptotic expansions of the leading bias and variance, (ii) asymptotic
Gaussian distributional approximation and valid statistical inference, (iii)
consistent standard error estimators, and (iv) consistent data-driven
bandwidth selection based on an asymptotic mean squared error (MSE) expansion.
All these results apply to both interior and boundary points in a fully
automatic and data-driven way, without requiring boundary-specific
transformations of the estimator or of the data, and without employing
additional tuning parameters (beyond the main bandwidth present in any
kernel-based nonparametric method).

As a substantive methodological application of our proposed density estimator,
we develop a novel discontinuity in density testing procedure. In a seminal
paper, \cite{McCrary_2008_JoE} proposed the idea of manipulation testing via
discontinuity in density testing for regression discontinuity (RD)\ designs,
and developed an implementation thereof using the density estimator of
\cite{Cheng-Fan-Marron_1997_AoS}, which requires pre-binning of the data and
choosing two tuning parameters. On the other hand, the new proposed
discontinuity in density test employing our density estimator only requires
the choice of one tuning parameter, and enjoys other features associated with
local polynomials methods. We also illustrate its performance with an
empirical application employing the canonical Head Start data in the context
of RD designs \citep{Ludwig-Miller_2007_QJE}. For introductions to RD designs,
and further references, see \cite{Imbens-Lemieux_2008_JoE},
\cite{Lee-Lemieux_2010_JEL}, and
\cite{Cattaneo-Titiunik-VazquezBare_2017_JPAM}. For recent papers on modern RD
methodology see, for example, \cite{Arai-Ichimura_2018_QE},
\cite{Ganong-Jager_2018_JASA}, \cite{Hyytinen-etal_2018_QE},
\cite{Dong-Lee-Gou_2019_wp}, and references therein.

Finally, we provide two general purpose software packages, for \texttt{Stata}
and \texttt{R}, implementing the main results discussed in the paper.
\cite{Cattaneo-Jansson-Ma_2018_Stata} discusses the package \texttt{rddensity}%
, which is specifically tailored to manipulation testing (i.e., two-sample
discontinuity in density testing), while
\cite{Cattaneo-Jansson-Ma_2019_lpdensity} discusses the package
\texttt{lpdensity}, which provides generic density estimation over the support
of the data.

The rest of the paper is organized as follows. Section
\ref{section:estimation} introduces the density estimator and Section
\ref{section:results} gives the main technical results. Section
\ref{section:rddensity} applies these results to nonparametric discontinuity
in density testing (i.e., manipulation testing), while Section
\ref{section:empapp} illustrates the new method with an empirical application.
Section \ref{section:conclusion} discusses extensions and concludes. The
supplemental appendix (SA hereafter) contains additional methodological and
technical results and reports all theoretical proofs. In addition, to conserve
space, we relegate to the SA and to our two companion software articles the
presentation of simulation evidence highlighting the finite sample properties
of our proposed density estimator.

\section{Boundary Adaptive Density Estimation\label{section:estimation}}

Suppose $\{x_{1},x_{2},\cdots,x_{n}\}$ is a random sample, where $x_{i}$ is a
continuous random variable with a smooth cumulative distribution function over
its support $\mathcal{X}\subseteq\mathbb{R}$. The probability density function
is $f(x)=\frac{\partial}{\partial x}\mathbb{P}[x_{i}\leq x]$, where the
derivative is interpreted as a one-sided derivative at a boundary point of
$\mathcal{X}$. Our results apply to bounded or unbounded support $\mathcal{X}%
$, which is an important feature in empirical applications employing density estimators.

Letting $\hat{F}(x)=\frac{1}{n}\sum_{i=1}^{n}%
%TCIMACRO{\TeXButton{\I}{\I}}%
%BeginExpansion
\I
%EndExpansion
(x_{i}\leq x)$ denote the classical empirical distribution function, our
proposed local polynomial density estimator is
\[
\hat{f}(x)=\mathbf{e}_{1}^{\prime}\mathbf{\hat{\beta}}(x),\qquad
\mathbf{\hat{\beta}}(x)=\underset{\mathbf{b}\in\mathbb{R}^{p+1}%
}{\operatorname{argmin}}\sum_{i=1}^{n}\left[  \hat{F}(x_{i})-\mathbf{r}%
_{p}(x_{i}-x)^{\prime}\mathbf{b}\right]  ^{2}K\left(  \frac{x_{i}-x}%
{h}\right)  ,
\]
where $\mathbf{e}_{1}=(0,1,0,\cdots,0)^{\prime}$ is the second $(p+1)$%
-dimensional unit vector, $\mathbf{r}_{p}(u)=(1,u,u^{2},\cdots,u^{p})^{\prime
}$ is a $p$-th order polynomial expansion, $K(\cdot)$ denotes a kernel
function, $h$ is a positive bandwidth, and $p\geq1$. In other words, we take
the empirical distribution function $\hat{F}$ as the starting point, then
construct a smooth local approximation to $\hat{F}$ using a polynomial
expansion, and finally obtain the density estimator $\hat{f}$ as the slope
coefficient in the local polynomial regression.

\begin{figure}[!tp]
	\begin{center}\caption{Graphical Illustration of Density Estimator.\label
{fig:lpdensity}}\vspace{-.1in}
		\includegraphics[width=0.8\textwidth]{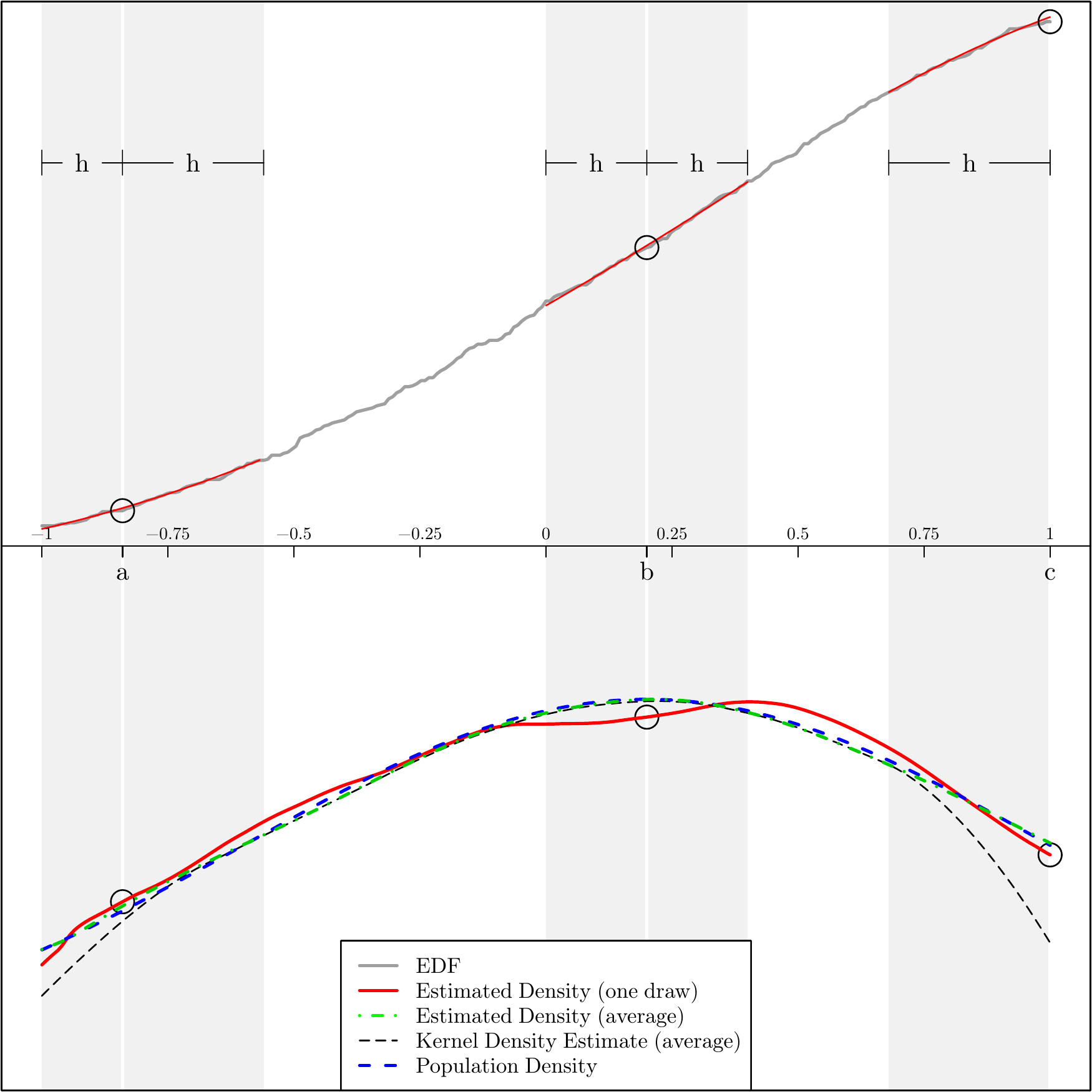}
	\end{center}
	\vspace{-0.05in}\footnotesize\textit{Notes}%
: (i) Constructed using companion \textsf{R} (and \textsf{Stata}%
) package described in \citet{Cattaneo-Jansson-Ma_2019_lpdensity}
with simulated data.
	\vspace{.5in}
\end{figure}

The idea behind the density estimator $\hat{f}(x)$ is explained graphically in
Figure \ref{fig:lpdensity}. In this figure, we consider three distinct
evaluation points on $\mathcal{X}=[-1,1]$: $a$ is near the lower boundary, $b
$ is an interior point, and $c=1$ is the upper boundary. The conventional
kernel density estimator, $\hat{f}_{\mathtt{KD}}(x)=\frac{1}{nh}\sum_{i=1}%
^{n}K\left(  \frac{x_{i}-x}{h}\right)  $, is valid for interior points, but
otherwise inconsistent. See, e.g., \cite{Wand-Jones_1995_Book} for a classical
reference. On the other hand, our density estimator $\hat{f}(x)$ is valid for
all evaluation points $x\in\mathcal{X}$ and can be used directly, without any
modifications to approximate the unknown density. Figure \ref{fig:lpdensity}
is constructed using $n=500$ observations. The top panel plots one realization
of the empirical distribution function $\hat{F}(x)$ in dark gray, and the
local polynomial fits for the three evaluation points $x\in\{a,b,c\}$ in red,
the latter implemented with $p=2$ (quadratic approximation) and bandwidth $h$
(different value for each evaluation point considered). The vertical light
gray areas highlight the localization region controlled by the bandwidth
choice, that is, only observations falling in these regions are used to smooth
out the empirical distribution function via local polynomial approximation,
depending on the evaluation point. The estimator $\hat{f}(x)$ is the slope
coefficient accompanying the first-order term in the local polynomial
approximation, which is depicted in the bottom panel of Figure
\ref{fig:lpdensity} as the solid line in red. The bottom panel also plots
three other curves: dashed blue line corresponding to the population density
function, dash-dotted green line corresponding to the average of our density
estimate over simulations, and dashed black line corresponding to the average
of the standard kernel density estimates $\hat{f}_{\mathtt{KD}}(x)$.

Figure \ref{fig:lpdensity} illustrates how our proposed density estimator
adapts to (near) boundary points automatically, showing graphically its good
performance in repeated samples. Evaluation point $b$ is an interior point
and, consequently, a symmetric smoothing around that point is employed, just
like the standard estimator $\hat{f}_{\mathtt{KD}}(x)$ does. On the other
hand, evaluation points $a$ and $c$ both exhibit boundary bias if the standard
kernel density estimator is used: point $a$ is near the boundary and hence
employs asymmetric smoothing, while point $c$ is at the upper boundary and
hence employs one-sided smoothing. In contrast, our proposed density estimator
$\hat{f}(x)$ automatically adapts to the boundary point, as the bottom panel
in Figure \ref{fig:lpdensity} illustrates.

\section{Main Technical Results\label{section:results}}

We summarize two main large sample results concerning the proposed density
estimator: (i) an asymptotic distributional approximation with precise leading
bias and variance characterizations, and (ii) a consistent standard error
estimator, which is also data-driven and fully automatic. Both results are
boundary adaptive and do not require prior knowledge of the shape of
$\mathcal{X}$. We report preliminary technical lemmas, additional theoretical
results, and detailed proofs in the SA to conserve space. Extensions and other
applications of our methods are mentioned in Section \ref{section:conclusion}.

\begin{Assumption}
[DGP]\label{ass:dgp} $\{x_{1},x_{2},\cdots,x_{n}\}$ is a random sample with
distribution function $F$ that is $p+1$ times continuously differentiable for
some $p\geq1$ in a neighborhood of the evaluation point $x$, and the
probability density function of $x_{i}$, denoted by $f$, is positive at $x$.
\end{Assumption}

This assumption imposes basic regularity conditions on the data generating
process, ensuring that $f(x)$ is well-defined and possesses enough smoothness.

\begin{Assumption}
[Kernel]\label{ass:kernel} The kernel function $K(\cdot)$ is nonnegative,
symmetric, and continuous on its support $[-1,1]$.
\end{Assumption}

This assumption is standard in nonparametric estimation, and is satisfied for
common kernel functions. We exclude kernels with unbounded support (e.g.,
Gaussian kernel) for simplicity, since such kernels will always hit
boundaries. Our results, however, can be extended to accommodate kernel
functions with unbounded support, albeit more cumbersome notation would be needed.

The following theorem gives a characterization of the asymptotic bias and
variance of $\hat{f}(x)$, as well as a valid distributional approximation. All
limits are taken as $n\rightarrow\infty$ (and $h\rightarrow0$) unless
explicitly stated otherwise, $\rightsquigarrow$ denotes weak convergence, and
$F^{(s)}(x)=\partial^{s}F(x)/\partial x^{s}$ denotes the derivative, or
one-sided derivative if at a boundary point, of $F(x).$

\begin{theorem}
[Distributional Approximation]\label{thm:AN} Suppose Assumption \ref{ass:dgp}
and \ref{ass:kernel} hold. If $nh^{2}\rightarrow\infty$ and $nh^{2p+1}=O(1)$,
then
\[
\frac{\hat{f}(x)-f(x)-h^{p}\mathcal{B}(x)}{\sqrt{\frac{1}{nh}\mathcal{V}(x)}%
}\rightsquigarrow\mathcal{N}(0,1),
\]
where, defining%
\begin{align*}
\mathbf{A}(x) &  =f(x)\int_{h^{-1}(\mathcal{X}-x)}\mathbf{r}_{p}%
(u)\mathbf{r}_{p}(u)^{\prime}K(u)\mathrm{d}u,\qquad\mathbf{a}(x)=f(x)\frac
{F^{(p+1)}(x)}{(p+1)!}\int_{h^{-1}(\mathcal{X}-x)}u^{p+1}\mathbf{r}%
_{p}(u)K(u)\mathrm{d}u,\\
\mathbf{B}(x) &  =f(x)^{3}\iint_{h^{-1}(\mathcal{X}-x)}\min\{u,v\}\mathbf{r}%
_{p}(u)\mathbf{r}_{p}(v)^{\prime}K(u)K(v)\text{d}u\mathrm{d}v,
\end{align*}
the asymptotic bias and variance are $\mathcal{B}(x)=\mathbf{e}_{1}^{\prime
}\mathbf{A}(x)^{-1}\mathbf{a}(x)$ and $\mathcal{V}(x)=\mathbf{e}_{1}^{\prime
}\mathbf{A}(x)^{-1}\mathbf{B}(x)\mathbf{A}(x)^{-1}\mathbf{e}_{1}$, respectively.
\end{theorem}

In this theorem, the integration region reflects the effect of boundaries.
Because $K(\cdot)$ is compactly supported, if $x$ is an interior point, we
have $h^{-1}(\mathcal{X}-x)\supset\lbrack-1,1]$ for $h$ small enough, thus
ensuring the kernel function is not truncated and the local approximation is
symmetric around $x$. On the other hand, for $x$ near or at a boundary of
$\mathcal{X}$ (i.e., for $h$ not small enough relative to the distance of $x$
to the boundary), we have $h^{-1}(\mathcal{X}-x)\not \supset \lbrack-1,1]$,
and the local approximation is asymmetric (or one-sided). It follows that the
density estimator $\hat{f}(x)$ is boundary adaptive and design adaptive, as in
the case of local polynomial regression \citep{Fan-Gijbels_1996_Book}.

A simple and automatic variance estimator is $\hat{\mathcal{V}}(x)=\mathbf{e}%
_{1}^{\prime}\mathbf{\hat{A}}(x)^{-1}\mathbf{\hat{B}}(x)\mathbf{\hat{A}%
}(x)^{-1}\mathbf{e}_{1}$, where%
\begin{align*}
\hat{\mathbf{A}}(x) &  =\frac{1}{nh}\sum_{i=1}^{n}\mathbf{r}_{p}\left(
\check{x}_{i}\right)  \mathbf{r}_{p}\left(  \check{x}_{i}\right)  ^{\prime
}K\left(  \check{x}_{i}\right) \\
\hat{\mathbf{B}}(x) &  =\frac{1}{n^{3}h^{3}}\sum_{i,j,k=1}^{n}\mathbf{r}%
_{p}\left(  \check{x}_{j}\right)  \mathbf{r}_{p}\left(  \check{x}_{k}\right)
^{\prime}K\left(  \check{x}_{j}\right)  K\left(  \check{x}_{k}\right)  \left[
%
%TCIMACRO{\TeXButton{\I}{\I}}%
%BeginExpansion
\I
%EndExpansion
(x_{i}\leq x_{j})-\hat{F}(x_{j})\right]  \left[
%TCIMACRO{\TeXButton{\I}{\I}}%
%BeginExpansion
\I
%EndExpansion
(x_{i}\leq x_{k})-\hat{F}(x_{k})\right]  ,
\end{align*}
with $\check{x}_{i}=h^{-1}(x_{i}-x)$ denoting the normalized observations to
save notation. Let $\rightarrow_{\mathbb{P}}$ denote convergence in probability.

\begin{theorem}
[Variance Estimation]\label{thm: variance estimation} If the conditions in
Theorem \ref{thm:AN} hold, then $\hat{\mathcal{V}}(x)\rightarrow_{\mathbb{P}%
}\mathcal{V}(x)$.
\end{theorem}

As shown in this theorem, the variance estimator $\hat{\mathcal{V}}(x)$ does
not require knowledge of the relative positioning of the evaluation point to
boundaries of $\mathcal{X}$, that is, $\hat{\mathcal{V}}(x)$ is also boundary
adaptive. A boundary adaptive bias estimator $\mathcal{\hat{B}}(x)$ can also
be constructed easily, as shown in the SA.

Using the results above, and under mild regularity conditions, it follows that
a pointwise approximate MSE-optimal bandwidth choice for our proposed density
estimator $\hat{f}(x)$ is
\[
h_{\mathtt{MSE}}(x)=\left(  \frac{\mathcal{V}(x)}{2p\mathcal{B}(x)^{2}%
}\right)  ^{1/(1+2p)}n^{-1/(1+2p)}\text{,}%
\]
which can be easily implemented by replacing $\mathcal{B}(x)$ and
$\mathcal{V}(x)$ with preliminary consistent estimators $\mathcal{\hat{B}}(x)$
and $\mathcal{\hat{V}}(x)$. The SA offers details on implementation and
consistency of this MSE-optimal bandwidth selector, which can be used to
establish its optimality in the sense of \cite{Li_1987_AoS}, and also
bandwidth selection for estimating higher-order density derivatives. We omit
these results here due to space limitations.

Finally, we recommend implementing the density estimator $\hat{f}(x)$ with
$p=2$, which corresponds to the minimal odd polynomial order choice (i.e.,
analogous to local linear regression). Higher-order local polynomials could be
used, but they typically exhibit erratic behavior near boundary points, and
lead to counter-intuitive weighting schemes. See \cite[Chapter 3.3]%
{Fan-Gijbels_1996_Book} for an automatic polynomial order selection methods
that can be applied to our estimator as well.

\section{Application to Manipulation Testing\label{section:rddensity}}

Testing for manipulation is useful when units are assigned to two (or more)
distinct groups using a hard-thresholding rule based on an observable
variable, as it provides an intuitive and simple method to check empirically
whether units are able to alter (i.e., manipulate) their assignment.
Manipulation tests are used in empirical work both as falsification tests of
regression discontinuity (RD) designs and as empirical tests with substantive
implications in other program evaluation settings. Available methods from the
RD literature include the original implementation of \cite{McCrary_2008_JoE}
based on \cite{Cheng-Fan-Marron_1997_AoS}, the empirical likelihood testing
procedure of \cite{Otsu-Xu-Matsushita_2014_JBES} based on boundary-corrected
kernels, and the finite sample binomial test presented in
\cite{Cattaneo-Titiunik-VazquezBare_2017_JPAM} based on local randomization ideas.

In this section, we introduce a new manipulation testing procedure based on
our proposed local polynomial density estimator. Our method requires choosing
only one tuning parameter, avoids pre-binning the data, and permits the use of
simple well-known weighting schemes (e.g., uniform or triangular kernel),
thereby avoiding the need of choosing the length and positions of bins for
pre-binning or employing more complicated boundary kernels. In addition, our
method is intuitive, easy-to-implement, and fully data-driven: bandwidth
selection methods are formally developed and implemented, along with valid
inference methods based on robust bias correction.

To describe the manipulation testing setup, suppose units are assigned to one
group (\textquotedblleft control\textquotedblright) if $x_{i}<\bar{x}$ and to
another group (\textquotedblleft treatment\textquotedblright) if $x_{i}%
\geq\bar{x}$. For example, in the application discussed below, we employ the
Head Start data, where $x_{i}$ is a poverty index at the county level,
$\bar{x}=59.1984$ is a fixed cutoff determining eligibility to the program.
The goal is to test formally whether the density $f(x)$ is continuous at
$\bar{x}$, using the two subsamples $\{x_{i}:x_{i}<\bar{x}\}$ and
$\{x_{i}:x_{i}\geq\bar{x}\}$, and thus the null and alternative hypotheses
are:
\[
\mathsf{H}_{0}:\lim_{x\uparrow\bar{x}}f(x)=\lim_{x\downarrow\bar{x}}%
f(x)\qquad\text{vs}\qquad\mathsf{H}_{1}:\lim_{x\uparrow\bar{x}}f(x)\neq
\lim_{x\downarrow\bar{x}}f(x).
\]

\begin{figure}[!tp]
\begin{center}\caption{Manipulation Testing, Head Start Data.\label
{fig:headstart}}
\vspace{-.1in}
\includegraphics[width=0.8\textwidth]{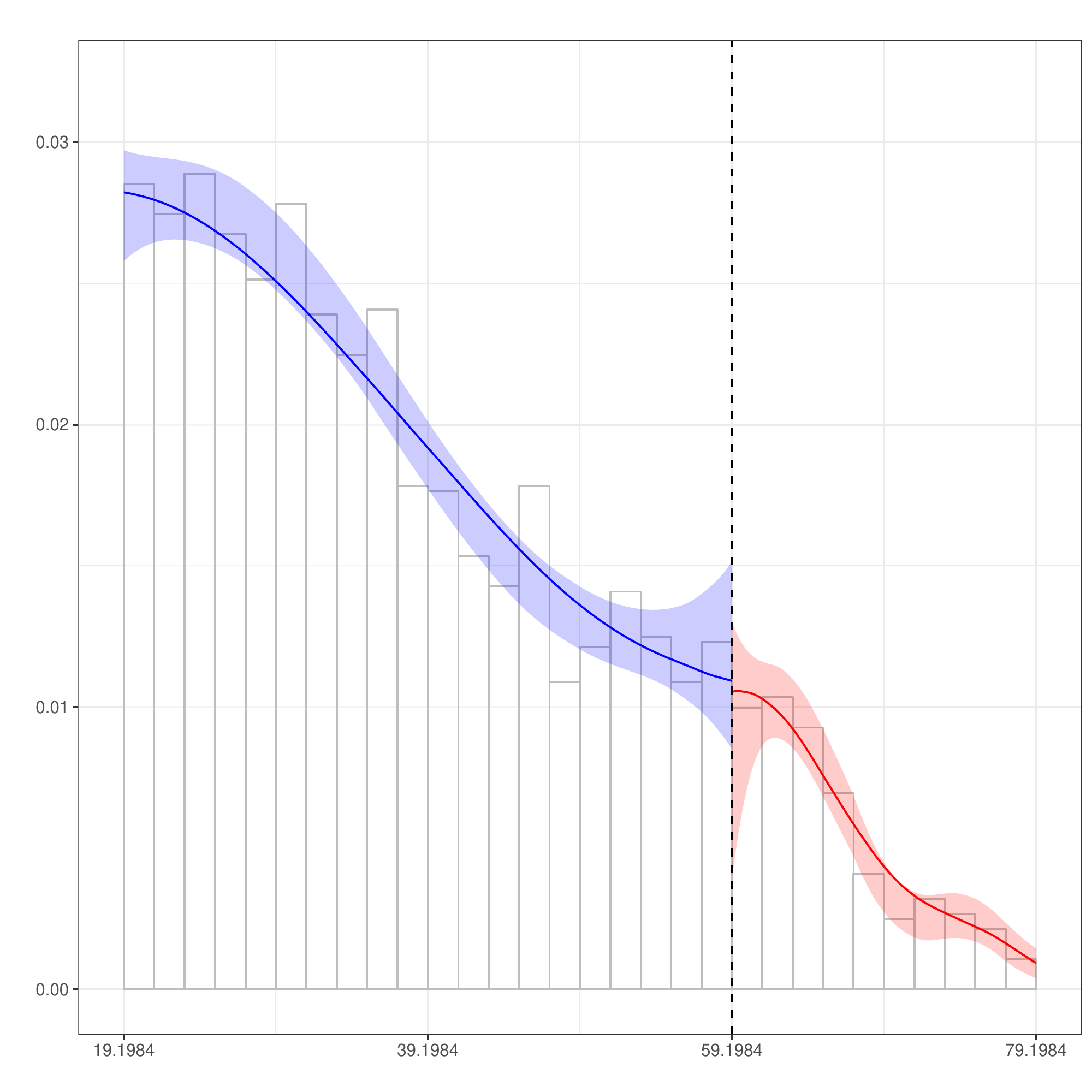}
\end{center}
\vspace{-0.05in}\footnotesize\textit{Notes}:
(i) Histogram estimate (light grey in background) of the running variable (poverty index) computed with default values in \texttt
{R};
(ii) local polynomial density estimate (solid blue and red) and robust bias corrected confidence intervals (shaded blue and red) computed using companion \texttt
{R} (and \texttt{Stata}) package described in \citet
{Cattaneo-Jansson-Ma_2018_Stata}; and
(ii) $n_{-}=2,504$, $n_{+}=300$, and $\bar{x}=59.1984$.
\vspace{.5in}
\end{figure}

This hypothesis testing problem induces a nonparametric boundary point at
$x=\bar{x}$ because two distinct densities need to be estimated, one from the
left and the other from the right. Our proposed density estimator $\hat{f}(x)
$ is readily applicable because it is boundary adaptive and fully automatic,
and it can also be used to plot the density near the cutoff in an automatic
way: see Figure \ref{fig:headstart} below for an example using the Head Start data.

Let $\hat{F}_{-}$ and $\hat{F}_{+}$ be the empirical distribution functions
constructed using only units with $x_{i}<\bar{x}$ and with $x_{i}\geq\bar{x}
$, respectively. Then, $\hat{f}$ can be applied twice, to the data below and
above the cutoff, to obtain two estimators of the density at the boundary
point $\bar{x}$, which we denote by $\hat{f}_{-}(\bar{x})$ and $\hat{f}%
_{+}(\bar{x})$, respectively. Thus, our proposed manipulation test statistic
takes the form:
\[
T_{p}(h)=\frac{\frac{n_{+}}{n}\hat{f}_{+}(\bar{x})-\frac{n_{-}}{n}\hat{f}%
_{-}(\bar{x})}{\sqrt{\frac{n_{+}}{n}\frac{1}{nh_{+}}\mathcal{\hat{V}}_{+}%
(\bar{x})+\frac{n_{-}}{n}\frac{1}{nh_{-}}\mathcal{\hat{V}}_{-}(\bar{x})}},
\]
where $n_{-}=\sum_{i=1}^{n}%
%TCIMACRO{\TeXButton{\I}{\I}}%
%BeginExpansion
\I
%EndExpansion
(x_{i}<x)$ and $n=n_{-}+n_{+}$, $\mathcal{\hat{V}}_{-}(x)$ and $\mathcal{\hat
{V}}_{+}(x)$ denote the variance estimators mentioned previously but now
computed for the two subsamples $x_{i}<\bar{x}$ and $x_{i}\geq\bar{x}$,
respectively, and $h_{-}$ and $h_{+}$ denote the bandwidths used below and
above $\bar{x}$. Employing our main theoretical results, we provide precise
conditions so that the finite sample distribution of $T_{p}(h)$ can be
approximated by the standard normal distribution, which leads to the following
result: under the regularity conditions given above, and if $n\min\{h_{-}%
^{2},h_{+}^{2}\}\rightarrow\infty$ and $n\max\{h_{-}^{1+2p},h_{+}%
^{1+2p}\}\rightarrow0$, then%
\begin{align*}
\text{Under }\mathsf{H}_{0}  & :\lim_{n\rightarrow\infty}\mathbb{P}%
[|T_{p}(h)|\geq\Phi_{1-\alpha/2}]=\alpha,\\
\text{Under }\mathsf{H}_{1}  & :\lim_{n\rightarrow\infty}\mathbb{P}%
[|T_{p}(h)|\geq\Phi_{1-\alpha/2}]=1,
\end{align*}
where $\Phi_{\alpha}$ denotes the $\alpha$-quantile of the standard Gaussian
distribution, $\alpha\in(0,1)$. This establishes asymptotic validity and
consistency of the $\alpha$-level testing procedure that rejects
$\mathsf{H}_{0}$ iff $|T(h)|\geq\Phi_{1-\alpha/2}$. The SA includes detailed
proofs, and related implementation details.

A key implementation issue of our manipulation test is the choice of bandwidth
$h$, a problem common to all nonparametric manipulation tests available in the
literature. To select $h$ in an automatic and data-driven way, we obtain an
approximate MSE-optimal bandwidth choice for the point estimator $\hat{f}%
_{+}(\bar{x})-\hat{f}_{-}(\bar{x})$, and then propose a consistent
implementation thereof, which is denoted by $\hat{h}_{p}$. We give the details
in the SA, where we also present alternative MSE-optimal bandwidth selectors
for each-side density estimator separately. Given the data-driven bandwidth
choice $\hat{h}_{p}$, or its theoretical (infeasible)\ counterpart $h_{p}$, we
propose a simple robust bias-corrected test statistic implementation following
ideas in \cite{Calonico-Cattaneo-Titiunik_2014_ECMA} and
\cite{Calonico-Cattaneo-Farrell_2018_JASA}; see the latter reference for
theoretical results on higher-order refinements and the important role of
pre-asymptotic variance estimation in the context of local polynomial
regression estimation. Specifically, our proposed data-driven robust
bias-corrected test statistic is $T_{p+1}(\hat{h}_{p})$, which rejects
$\mathsf{H}_{0}$ iff $|T_{p+1}(\hat{h}_{p})|\geq\Phi_{1-\alpha/2}$ for a
nominal $\alpha$-level test. This approach corresponds to a special case of
manual bias-correction together with the corresponding adjustment of
Studentization. A natural choice is $p=2$, and this is the default in our
companion \texttt{Stata} and \texttt{R} software implementations.

\section{Empirical Illustration\label{section:empapp}}

We apply our manipulation test to the data of \cite{Ludwig-Miller_2007_QJE} on
the original Head Start implementation in the U.S. In this empirical
application, a discontinuity on access to program funds at the county level
occurred in $1965$ when the program was first implemented: the federal
government provided grant writing assistance to the $300$ poorest counties as
measured by a poverty index, which was computed in $1965$ using $1960$ Census
variables, thus creating a discontinuity in program elegibility. Using our
notation, $x_{i}$ denotes the poverty index for county $i$, and $\bar
{x}=59.1984$ is the cutoff point (i.e., the poverty index of the $300$-th
poorest municipality).

A manipulation test in this context amounts to testing whether there is a
disproportional number of counties are situated above $\bar{x}$ relative to
those present below the cutoff. Figure \ref{fig:headstart} presents the
histogram of counties below and above the cutoff together with our local
polynomial density estimate and associated pointwise robust bias-corrected
confidence intervals over a grid of points near the cutoff $\bar{x}$,
implemented using $p=2$ and the MSE-optimal data-driven bandwidth estimate.
Table \ref{table:headstart} presents the empirical results from our
manipulation test. We consider two main approaches, both covered by our
theoretical work and available in our software implementation: (i) using two
distinct bandwidths on each side of the cutoff ($h_{-}\neq h_{+}$), and (ii)
using a common bandwidth for each side of the cutoff ($h_{-}=h_{+}$), with
$h_{-}$ and $h_{+}$ denoting the bandwidth on the left and on the right,
respectively. For each case, we consider three distinct implementations of our
manipulation test, which varies the degree of polynomial approximation used to
smooth out the empirical distribution function: $T_{q}(h_{p})$ denotes the
test statistic constructed using a $q$-th order local polynomial density
estimator, with bandwidth choice that is MSE-optimal for $p$-th order local
polynomial density estimator. For example, our recommended choice is
$T_{3}(h_{2})$, with either common bandwidth or two different bandwidths,
which amounts to first choosing MSE-optimal bandwidth(s) for a local quadratic
fit, and then conducting inference using a cubic approximation. This approach
is the simplest implementation of the robust bias correction inference:
$T_{p}(h_{p})$ does not lead to a valid inference approach because a
first-order bias will make the test over-reject the null hypothesis. We also
report the original implementation of the McCrary test for comparison.

Our empirical results show no evidence of manipulation. In fact, this finding
is consistent with the underlying institutional knowledge of the program: the
poverty index was constructed in $1965$ at the federal level using county
level information from the $1960$ Census, which implies it is indeed highly
implausible that individual counties could have manipulated their assigned
poverty index. Our findings are robust to different bandwidth and local
polynomial order specifications. Finally, we note two theory-based empirical
findings: (i) our proposed manipulation test employs robust bias-corrected
methods, and hence leads to asymmetric confidence intervals (not necessarily
centered around the density point estimator); and (ii)\ the effective sample
size of the original McCrary test is much smaller than our proposed
manipulation test because of the pre-binning of the data, and hence can lead
to important reduction in power of the test.

\section{Conclusion\label{section:conclusion}}

We introduced a boundary adaptive kernel-based density estimator employing
local polynomial methods, which requires choosing only one tuning parameter
and does not require boundary-specific data transformations (such as
pre-binning). We studied the main asymptotic properties of the estimator, and
used these results to developed a new manipulation test via discontinuity in
density testing. Several extensions and generalizations of our results are
underway in ongoing work, and two distinct general purpose software packages
in \texttt{Stata} and \texttt{R} are readily available \citet{Cattaneo-Jansson-Ma_2018_Stata,Cattaneo-Jansson-Ma_2019_lpdensity}.

\newpage%

%TCIMACRO{\TeXButton{\singlespacing}{\onehalfspacing\singlespacing}}%
%BeginExpansion
\onehalfspacing\singlespacing
%EndExpansion

\bibliographystyle{jasa}
\bibliography{Cattaneo-Jansson-Ma_2019_JASA}

\newpage%

%TCIMACRO{\TeXButton{Figure: heuristics}{\begin{figure}[!tp]
%	\begin{center}\caption{Graphical Illustration of Density Estimator.\label
%{fig:lpdensity}}\vspace{-.1in}
%		\includegraphics[width=0.8\textwidth]{\pathpaper
%/simulation/figures/lpdensity.pdf}
%	\end{center}
%	\vspace{-0.05in}\footnotesize\textit{Notes}%
%: (i) Constructed using companion \textsf{R} (and \textsf{Stata}%
%) package described in \citet{Cattaneo-Jansson-Ma_2019_lpdensity}
%with simulated data.
%	\vspace{.5in}
%\end{figure}}}%
%BeginExpansion
%EndExpansion
%

%TCIMACRO{\TeXButton{Figure: empapp}{\begin{figure}[!tp]
%\begin{center}\caption{Manipulation Testing, Head Start Data.\label
%{fig:headstart}}
%\vspace{-.1in}
%\includegraphics[width=0.8\textwidth]{\pathpaper/empapp/output/headstart.pdf}
%\end{center}
%\vspace{-0.05in}\footnotesize\textit{Notes}:
%(i) Histogram estimate (light grey in background) of the running variable (poverty index) computed with default values in \texttt
%{R};
%(ii) local polynomial density estimate (solid blue and red) and robust bias corrected confidence intervals (shaded blue and red) computed using companion \texttt
%{R} (and \texttt{Stata}) package described in \citet
%{Cattaneo-Jansson-Ma_2018_Stata}; and
%(ii) $n_{-}=2,504$, $n_{+}=300$, and $\bar{x}=59.1984$.
%\vspace{.5in}
%\end{figure}}}%
%BeginExpansion
\renewcommand{\arraystretch}{1.2}
\begin{table}[!tp]\vspace{-.25in}
\begin{center}\caption{Manipulation Testing, Head Start Data.\label
{table:headstart}}\vspace{-.1in}\footnotesize\resizebox{\textwidth}{!}%
{%latex.default(round(Result[, c(7:8, 1:6)], 3), file = "output/headstart.txt",     append = FALSE, table.env = FALSE, center = "none", title = "",     n.cgroup = c(2, 2, 2, 2), cgroup = c("Pre-binning", "Bandwidths",         "Eff. $n$", "Test"), colheads = c("left", "right", "left",         "right", "left", "right", "$T$", "$p$-val"), n.rgroup = c(3,         3, 1), rgroup = c("$h_-\\neq h_+$", "$h_-= h_+$", "McCrary"),     rowname = c("$T_2(\\hat{h}_1)$", "$T_3(\\hat{h}_2)$", "$T_4(\\hat{h}_3)$",         "$T_2(\\hat{h}_1)$", "$T_3(\\hat{h}_2)$", "$T_4(\\hat{h}_3)$",         ""))%
\begin{tabular}{lrrcrrcrrcrr}
\hline\hline
\multicolumn{1}{l}{\bfseries }&\multicolumn{2}{c}{\bfseries Pre-binning}&\multicolumn{1}{c}{\bfseries }&\multicolumn{2}{c}{\bfseries Bandwidths}&\multicolumn{1}{c}{\bfseries }&\multicolumn{2}{c}{\bfseries Eff. $n$}&\multicolumn{1}{c}{\bfseries }&\multicolumn{2}{c}{\bfseries Test}\tabularnewline
\cline{2-3} \cline{5-6} \cline{8-9} \cline{11-12}
\multicolumn{1}{l}{}&\multicolumn{1}{c}{left}&\multicolumn{1}{c}{right}&\multicolumn{1}{c}{}&\multicolumn{1}{c}{left}&\multicolumn{1}{c}{right}&\multicolumn{1}{c}{}&\multicolumn{1}{c}{left}&\multicolumn{1}{c}{right}&\multicolumn{1}{c}{}&\multicolumn{1}{c}{$T$}&\multicolumn{1}{c}{$p$-val}\tabularnewline
\hline
{\bfseries $h_-\neq h_+$}&&&&&&&&&&&\tabularnewline
~~$T_2(\hat{h}_1)$&$$&$$&&$15.771$&$ 2.326$&&$ 581$&$ 65$&&$ 0.024$&$0.981$\tabularnewline
~~$T_3(\hat{h}_2)$&$$&$$&&$19.776$&$ 8.296$&&$ 762$&$210$&&$-1.146$&$0.252$\tabularnewline
~~$T_4(\hat{h}_3)$&$$&$$&&$32.487$&$10.808$&&$1598$&$232$&&$-1.083$&$0.279$\tabularnewline
\hline
{\bfseries $h_-= h_+$}&&&&&&&&&&&\tabularnewline
~~$T_2(\hat{h}_1)$&$$&$$&&$ 3.274$&$ 3.274$&&$  99$&$ 95$&&$-1.355$&$0.175$\tabularnewline
~~$T_3(\hat{h}_2)$&$$&$$&&$ 9.213$&$ 9.213$&&$ 316$&$221$&&$-0.515$&$0.607$\tabularnewline
~~$T_4(\hat{h}_3)$&$$&$$&&$12.270$&$12.270$&&$ 419$&$243$&&$-0.712$&$0.477$\tabularnewline
\hline
McCrary&$76$&$60$&&$13.950$&$13.950$&&$  24$&$ 24$&&$ 0.142$&$0.887$\tabularnewline
\hline
\end{tabular}
}
\end{center}
\medskip\footnotesize\textit{Notes}:
(i) $T_p(h)$ denotes the manipulation test statistic using $p$-th order density estimators with bandwidth choice $h$ (which could be common on both sides or different on either side of the cutoff), and $\hat
{h}%
_p$ denotes the estimated MSE-optimal bandwidths for $p$-th order density estimator or difference of estimators (depending on the case considered);
(ii) Columns under ``Bandwidths'' report estimated MSE-optimal bandwidths, Columns under ``Eff. $n$'' report effective sample size on either side of the cutoff, and Columns under ``Test'' report value of test statistic ($T$) and two-sided p-value ($p$-val);
(iii) first three rows allow for different bandwidths on each side of the cutoff, while the next three rows employ a common bandwidth on both sides of the cutoff (chosen to be MSE-optimal for the difference of density estimates). All estimates are obtained using companion \texttt
{R} (and \texttt{Stata}) package described in \citet
{Cattaneo-Jansson-Ma_2018_Stata}; and
(iv) the last row, labeled ``McCrary'', corresponds to the original implementation of \citet
{McCrary_2008_JoE}%
, and therefore columns under ``Pre-binning'' report the total number of bins used for pre-bining of the data and columns under ``Eff. $n$'' report the number of bins used for local linear density estimation.
\end{table}%
%EndExpansion

\end{document}

% % % % % % % % % % % % % % Title Page % % % % % % % % % % % % % %
\title{\vspace{-0.25in} 
 Simple Local Polynomial Density Estimators \\
 Supplemental Appendix\bigskip\bigskip}
\author{Matias D. Cattaneo\thanks{Department of Operations Research and Financial Engineering, Princeton University.} \and 
	    Michael Jansson\thanks{Department of Economics, UC Berkeley and \emph{CREATES}.} \and 
        Xinwei Ma\thanks{Department of Economics, UC San Diego.}
}\maketitle

\vspace{0.2in}\begin{abstract}
This Supplemental Appendix contains general theoretical results and their proofs, which encompass those discussed in the main paper, discusses additional methodological and technical results, and reports simulation evidence.
\end{abstract}

\thispagestyle{empty}

\clearpage

% % % % % % % % % % % % % % TOC % % % % % % % % % % % % % %
\setcounter{tocdepth}{2} 
\tableofcontents 
\thispagestyle{empty}
\clearpage

\onehalfspacing
\setcounter{page}{1}
\pagestyle{plain}

% % % % % % % % % % % % % % Beginning of the Document % % % % % % % % % % % % % %
\pagestyle{plain}

\section{Setup}

We repeat the setup in the main paper for completeness. Recall that $\{ x_i \}_{1\leq i\leq n}$ is a random sample from the cumulative distribution function (hereafter CDF) $F$, supported on $\mathcal{X}=[\xmin,\xmax]$. Note that it is possible to have $\xmin=-\infty$ and/or $\xmax=\infty$. We will assume both $\xmin$ and $\xmax$ are finite, to facilitate discussion on boundary estimation issues. 

Define the empirical distribution function (hereafter EDF) 
\begin{align*}
\tilde{F}(x) &= \frac{1}{n}\sum_i \Indicator[x_i\leq x]. 
\end{align*}
Note that in the main paper, we use $\hat{F}(\cdot)$ to denote the above EDF We avoid such notation in this Supplemental Appendix, and instead use $\tilde{F}(\cdot)$, because a (smoothed) CDF estimator can be obtained from our local polynomial approach.

Given $p\in\mathbb{N}$, our local polynomial distribution estimator is defined as
\begin{align*}
\hat{\bbeta}_p(x) &= \arg\min_{\mathbf{b}\in\mathbb{R}^{p+1}} \sum_i \left( \tilde{F}(x_i) - \br_p(x_i-x)^\Trans\mathbf{b} \right)^2 K\left(\frac{x_i-x}{h}\right),
\end{align*} 
where $\br_p(u)=[1, u, u^2, \cdots, u^p]$ is a (one-dimensional) polynomial expansion; $K$ is a kernel function whose properties are to be specified later; $h=h_n$ is a bandwidth sequence. The estimator, $\hat{\bbeta}_p(x)$, is motivated as a local Taylor series expansion, hence the target parameter is (i.e., the population counterpart, assuming exists)
\begin{align*}
\bbeta_p(x) &= \left[ \frac{1}{0!}F(x),\ \frac{1}{1!}F^{(1)}(x),\ \cdots,\ \frac{1}{p!}F^{(p)}(x) \right]^\Trans.
\end{align*} 
Therefore, we also write
\begin{align*}
\hat{\bbeta}_p(x) &= \left[ \frac{1}{0!}\hat{F}_p(x),\ \frac{1}{1!}\hat{F}_p^{(1)}(x),\ \cdots,\ \frac{1}{p!}\hat{F}_p^{(p)}(x) \right]^\Trans,
\end{align*}
or equivalently, $\hat{F}_p^{(v)} = v! \be_v^\Trans \hat{\bbeta}_p(x)$, provided that $v\leq p$, and $\be_v$ is the $(v+1)$-th unit vector of $\mathbb{R}^{p+1}$. (The subscript $p$ is omitted in the main paper to economize notation.) We also use $f=F^{(1)}$ to denote the corresponding probability density function (hereafter PDF) for convenience. 

The estimator has the following matrix form:
\begin{align*}
\hat{\bbeta}_p(x) &= \bH^{-1} \left( \frac{1}{n} \bX_h^\Trans \bK_h \bX_h \right)^{-1} \left(\frac{1}{n} \bX_h^\Trans \bK_h \bY \right),\qquad \bX_h = \Big[ \left(\frac{x_i-x}{h}\right)^{j}  \Big]_{1\leq i\leq n,\ 0\leq j\leq p}
\end{align*}
where 
$\bK_h$ is a diagonal matrix collecting $\{ h^{-1}K((x_i-x)/h) \}_{1\leq i\leq n}$, and $\bY$ is a column vector collecting $\{ \tilde{F}(x_i) \}_{1\leq i\leq n}$. We also adopt the convention $K_h(u) = h^{-1}K(u/h)$. 

In this Supplemental Appendix, we use $n$ to denote sample size, and limits are taken with $n\to \infty$, unless otherwise specified. The standard Euclidean norm is denoted by $|\cdot|$, and other norms will be defined at their first appearances. Maximum and minimum of two real numbers $a$ and $b$ are denoted by $a\vee b$ and $a\wedge b$, respectively. For sequence of numbers (or random variables), $a_n\precsim b_n$ implies $\limsup_n |a_n/b_n|$ is finite, and $a_n\asymp b_n$ implies both directions. The notation $a_n\precsim_\Prob b_n$ is used to denote that $|a_n/b_n|$ is asymptotically tight: $\limsup_{\epsilon\uparrow \infty}\limsup_{n}\Prob[|a_n/b_n|\geq \epsilon]=0$. $a_n\asymp_{\Prob} b_n$ implies both $a_n\precsim_\Prob b_n$ and  $b_n\precsim_\Prob a_n$. When $b_n$ is a sequence of nonnegative numbers, $a_n=O(b_n)$ is sometimes used for $a_n\precsim b_n$, so does $a_n=\Op(b_n)$. For probabilistic convergence, we use $\to_\Prob$ for convergence in probability and $\rightsquigarrow$ for weak convergence (convergence in distribution). Standard normal distribution is denoted as $\mathcal{N}(0,1)$, with CDF $\Phi$ and PDF $\phi$. Throughout, we use $C$ to denote generic constants which do not depend on sample size. The exact value can change given the context.

\subsection{Overview of Main Results}

In this subsection, we give an overview of our results, including a (first order) mean squared error (hereafter MSE) expansion, and asymptotic normality. Fix some $v\geq 1$ and $p$, we have the following:
\begin{align*}
\left|\hat{F}_p^{(v)}(x) - F^{(v)}(x)\right| &= \Op\left(h^{p+1-v} \mathcal{B}_{p,v}(x) + h^{p+2-v} \tilde{\mathcal{B}}_{p,v}(x) + \sqrt{\frac{1}{nh^{2v-1}}\mathcal{V}_{p,v}(x)} \right).
\end{align*}
The previous result gives MSE expansion for derivative estimators, $1\leq v\leq p$, but not for $v=0$. With $v=0$, $\hat{F}_p(x)$ is essentially a smoothed EDF, which estimates the CDF $F(x)$. Since $F(x)$ is $\sqrt{n}$-estimable, one should be expected that the estimated distribution function will have very different properties compared to the estimated derivatives. Indeed, we have
\begin{align*}
\left|\hat{F}_p(x) - F(x)\right| &= \Op \left(h^{p+1} \mathcal{B}_{p,0}(x) + h^{p+2} \tilde{\mathcal{B}}_{p,0}(x)  + \sqrt{\frac{1}{n}\mathcal{V}_{p,0}(x)} \right).
\end{align*}
There is another complication, however, when $x$ is in the boundary region. For a drifting sequence $x$ in the boundary region, the EDF $\tilde{F}(x)$ is ``super-consistent'' in the sense that it converges at rate $\sqrt{h/n}$. The reason is that when $x$ is near $\xmin$ or $\xmax$, $\tilde{F}(x)$ is essentially estimating 0 or 1, and the variance, $F(x)(1-F(x))$ vanishes asymptotically, giving rise to the additional factor $\sqrt{h}$. This is shared by our estimator: for $v=0$ and $x$ in the boundary region, the CDF estimator $\tilde{F}_p(x)$ is super-consistent, with $\mathcal{V}_{p,0}(x)\asymp h$. 

Also note that for the MSE expansion, we provide not only the first order bias but also the second order bias. The second order bias will be used for bandwidth selection, since it is well-known that in some cases the first order bias can vanish. (More precisely, when $x$ is an interior evaluation point and $p-v$ is even. See, for example, \citealt{Fan-Gijbels_1996_Book}.) 

The MSE expansion provides the rate of convergence of our estimator. The following shows that, under suitable regularity conditions, they are also asymptotically normal. Again first consider $v\geq 1$.
\begin{align*}
\sqrt{nh^{2v-1}}\Big( \hat{F}_p^{(v)}(x) - F^{(v)}(x) - h^{p+1-v} \mathcal{B}_{p,v}(x) \Big) \rightsquigarrow \mathcal{N}\Big(0, \mathcal{V}_{p,v}(x)\Big),
\end{align*} 
provided that the bandwidth is not too large, so that after scaling, the remaining bias does not feature in first-order asymptotics. For $v=0$, i.e. the smoothed EDF, we have
\begin{align*}
\sqrt{\frac{n}{\mathcal{V}_{p,0}(x)}}\Big( \hat{F}_p(x) - F(x) - h^{p+1} \mathcal{B}_{p,0}(x) \Big) \rightsquigarrow \mathcal{N}\Big(0, 1\Big),
\end{align*}
where we moved the variance $\mathcal{V}_{p,0}(x)$ as a scaling factor in the above display, to encompass the situation where $x$ lies near boundaries. 

\subsection{Some Matrices}

In this subsection we collect some matrices which will be used throughout this Supplemental Appendix. They show up in asymptotic results as components of the (leading) bias and variance. Note that $x$ can be either a fixed point, or it can be a drifting sequence to capture the issue of estimation and inference in boundary regions. For the latter case, $x$ takes the form $x=\xmin+ch$ or $x=\xmax-ch$ for some $c\in[0,1)$. 

Define
\begin{align*}
&\ \bS_{p,x} = \int_{\xminh}^{\xmaxh} \br_p(u)\br_p(u)^\Trans K(u)\diff u,\quad \bc_{p,x} = \int_{\xminh}^{\xmaxh} \br_p(u)u^{p+1} K(u)\diff u,\quad \tilde{\bc}_{p,x} = \int_{\xminh}^{\xmaxh} \br_p(u)u^{p+2} K(u)\diff u,\\
&\ \bGamma_{p,x} = \iint_{\xminh}^{\xmaxh} (u\wedge v) \br_p(u)\br_p(v) K(u)K(v)\diff u\diff v,\quad \bT_{p,x} = \int_{\xminh}^{\xmaxh} \br_p(u)\br_p(u)^\Trans K(u)^2\diff u.
\end{align*}
Later we will assume that the kernel function $K$ is supported on $[-1, 1]$, hence with a shrinking bandwidth sequence $h\downarrow 0$, the region of integration in the above display can be replaced by
\begin{center}
\begin{tabular}{ccc}
\hline 
$x$ & $(\xmin-x)/h$ & $(\xmax-x)/h$\\  \hline
$x$ interior & $-1$  & $+1$ \\
$x=\xmin+ch$ in lower boundary & $-c$ & $+1$\\
$x=\xmax-ch$ in upper boundary & $-1$ & $+c$\\ \hline
\end{tabular}
\end{center} 
Since we do not allow $\xmin=\xmax$, no drifting sequence $x$ can be in both lower and upper boundary regions, at least in large samples.  

\section{Large Sample Properties}

\subsection{Assumptions}

In this section we give assumptions, preliminary lemmas and our main results. Other assumptions specific to certain results will be given in corresponding sections. 

Let $\mathcal{O}$ be a connected subset of $\mathbb{R}$ with nonempty interior, $\mathcal{C}^s(\mathcal{O})$ denotes functions that are at least $s$-times continuously differentiable in the interior of $\mathcal{O}$, and that the derivatives can be continuously extended to the boundary of $\mathcal{O}$. 

\begin{Assumption}[DGP]
\label{assumption: 1 dgp}$ $\\
$\{x_i\}_{1\leq i\leq n}$ is a random sample from distribution $F$, supported on $\mathcal{X}=[\xmin,\xmax]$. Further, $F\in\mathcal{C}^{\alpha_x}(\mathcal{X})$ for some $\alpha_x\geq 1$, and $f(x)=F^{(1)}(x)>0$ for all $x\in\mathcal{X}$. 
\end{Assumption} 

\begin{Assumption}[Kernel]
\label{assumption: 2 kernel}$ $\\
The kernel function $K(\cdot)$ is nonnegative, symmetric, and belongs to $\mathcal{C}^0([-1,1])$. Further, it integrates to one: $\int_{\mathbb{R}}K(u)\diff u=1$. 
\end{Assumption}

\subsection{Preliminary Lemmas}

We first consider the object $\bX_h^{\Trans}\bK_h\bX_h/n$
\begin{lem}\label{lem1: denominator}
Assume Assumptions \ref{assumption: 1 dgp} and \ref{assumption: 2 kernel} hold,  $h\to 0$ and $nh\to \infty$. Then
\begin{align*}
\frac{1}{n}\bX_h^\Trans\bK_h\bX_h = f(x)\bS_{p,x} + o(1) + \Op\left( 1/\sqrt{nh} \right).
\end{align*} 
\end{lem} 
Lemma \ref{lem1: denominator} shows that the matrix $\bX_h^{\Trans}\bK_h\bX_h/n$ is asymptotically invertible. Also note that this result covers both interior and boundary evaluation point $x$, and depending on the nature of $x$, the exact form of $\bS_{p,x}$ differs. 

With simple algebra, one has
\begin{align*}
\hat{\bbeta}_p(x) - \bbeta_p(x) &= \bH^{-1} \left( \frac{1}{n} \bX_h^\Trans \bK_h \bX_h \right)^{-1} \left(\frac{1}{n} \bX_h^\Trans \bK_h (\bY - \bX\bbeta_p(x)) \right),
\end{align*}
and the following gives a further decomposition of the ``numerator.''
\begin{align*}
&\ \frac{1}{n} \bX_h^\Trans \bK_h (\bY - \bX\bbeta_p(x)) = \frac{1}{n}\sum_i \br_p\left(\frac{x_i-x}{h}\right)\Big(\tilde{F}(x_i) - \br_p(x_i-x)^\Trans \bbeta_p(x)\Big)K_h(x_i-x) \\
&\qquad\qquad=\ \frac{1}{n}\sum_i \br_p\left(\frac{x_i-x}{h}\right)\Big(F(x_i) - \br_p(x_i-x)^\Trans \bbeta_p(x)\Big)K_h(x_i-x) \\
&\qquad\qquad+ \int_{\xminh}^{\xmaxh} \br_p(u)\Big(\tilde{F}(x+hu) - F(x+hu)\Big)K(u)f(x+hu)\diff u\\
&\qquad\qquad+ \frac{1}{n}\sum_i \br_p\left(\frac{x_i-x}{h}\right)\Big(\tilde{F}(x_i) - F(x_i)\Big)K_h(x_i-x) - \int_{\xminh}^{\xmaxh} \br_p(u)\Big(\tilde{F}(x+hu) - F(x+hu)\Big)K(u)f(x+hu)\diff u.
\end{align*}
The first part represents the smoothing bias, and the second part can be analyzed as a sample average. The real challenge comes from the third term, which can have a nonnegligible (first order) contribution. We further decompose it as
\begin{align*}
&\ \frac{1}{n}\sum_i \br_p\left(\frac{x_i-x}{h}\right)\Big(\tilde{F}(x_i) - F(x_i)\Big)K_h(x_i-x) = \frac{1}{n^2}\sum_{i,j} \br_p\left(\frac{x_i-x}{h}\right)\Big(\Indicator[x_j\leq x_i] - F(x_i)\Big)K_h(x_i-x)\\
=&\ \frac{1}{n^2}\sum_i \br_p\left(\frac{x_i-x}{h}\right)\Big(1 - F(x_i)\Big)K_h(x_i-x) 
+ \frac{1}{n^2}\sum_{i,j;i\neq j} \br_p\left(\frac{x_i-x}{h}\right)\Big(\Indicator[x_j\leq x_i] - F(x_i)\Big)K_h(x_i-x).
\end{align*}
As a result,
\begin{align*}
&\ \frac{1}{n}\sum_i \br_p\left(\frac{x_i-x}{h}\right)\Big(\tilde{F}(x_i) - \br_p(x_i-x)^\Trans \bbeta_p(x)\Big)K_h(x_i-x) \\
\tag{smoothing bias $\hat{\bB}_{\mathtt{S}}$}
&\qquad\qquad=\ \frac{1}{n}\sum_i \br_p\left(\frac{x_i-x}{h}\right)\Big(F(x_i) - \br_p(x_i-x)^\Trans \bbeta_p(x)\Big)K_h(x_i-x) \\
\tag{linear variance $\hat{\bL}$}
&\qquad\qquad+ \int_{\xminh}^{\xmaxh} \br_p(u)\Big(\tilde{F}(x+hu) - F(x+hu)\Big)K(u)f(x+hu)\diff u\\
\tag{leave-in bias $\hat{\bB}_{\mathtt{LI}}$}
&\qquad\qquad+ \frac{1}{n^2}\sum_i \br_p\left(\frac{x_i-x}{h}\right)\Big(1 - F(x_i)\Big)K_h(x_i-x)\\
&\qquad\qquad+ \frac{1}{n^2}\sum_{i,j;i\neq j} \Big\{  \br_p\left(\frac{x_i-x}{h}\right)\Big(\Indicator[x_j\leq x_i] - F(x_i)\Big)K_h(x_i-x)\\
\tag{quadratic variance $\hat{\bR}$}
&\qquad\qquad\qquad - \Expectation\left[ \br_p\left(\frac{x_i-x}{h}\right)\Big(\Indicator[x_j\leq x_i] - F(x_i)\Big)K_h(x_i-x)\Big| x_j \right]\Big\}.
\end{align*}
To provide intuition for the above decomposition, the smoothing bias is a typical feature of nonparametric estimators; leave-in bias arises since each observation is used twice, in constructing the EDF $\tilde{F}$ and as a design point (that is, $\tilde{F}$ has to be evaluated at $x_i$); and a second order U-statistic shows up because the ``dependent variable,'' $\bY$, is estimated, which leads to double summation. 

We first analyze the bias terms. 
\begin{lem}\label{lem2: smoothing bias}
Assume Assumptions \ref{assumption: 1 dgp} and \ref{assumption: 2 kernel} hold with $\alpha_x\geq p+1$, $h\to 0$ and $nh\to\infty$. Then
\begin{align*}
\hat{\bB}_{\mathtt{S}} &= h^{p+1}\frac{F^{(p+1)}(x)f(x)}{(p+1)!}\bc_{p,x} + \op(h^{p+1}),\qquad \hat{\bB}_{\mathtt{LI}} = \Op\left(n^{-1}\right).
\end{align*}
\end{lem}

By imposing additional smoothness, it is also possible to characterize the next term in the smoothing bias, which has order $h^{p+2}$. We report the higher order bias in a later section as it is used for bandwidth selection. 

Next we consider the ``influence function'' part, $\hat{\bL}$. This term is crucial in the sense that (under suitable conditions so that $\hat{\bR}$ becomes negligible) it determines the asymptotic variance of our estimator, and with correct scaling, it is asymptotically normally distributed. 
\begin{lem}\label{lem3: linear variance}
Assume Assumptions \ref{assumption: 1 dgp}and \ref{assumption: 2 kernel} hold with $\alpha_x\geq 2$, $h\to 0$ and $nh\to\infty$. Define the scaling matrix 
\begin{align*}
\bN_x &= \begin{cases}
\mathrm{diag}\Big\{ 1,\qquad h^{-1/2},\ h^{-1/2},\ \cdots,\ h^{-1/2}  \Big\} & \text{$x$ interior},\\
\mathrm{diag}\Big\{ h^{-1/2},\ h^{-1/2},\ h^{-1/2},\ \cdots,\ h^{-1/2}  \Big\} & \text{$x$ boundary},\\
\end{cases}
\end{align*}
then
\begin{align*}
\sqrt{n}\bN_x \Big[f(x)\bS_{p,x}\Big]^{-1}\hat{\bL} \rightsquigarrow \mathcal{N}(\mathbf{0},\ \mathsf{V}_{p,x}),
\end{align*}
with
\begin{align*}
\mathsf{V}_{p,x} = \begin{cases}
F(x)(1-F(x))\be_0\be_0^\Trans  + f(x) (\bI-\be_0\be_0^\Trans)\bS_{p,x}^{-1}\bGamma_{p,x}\bS_{p,x}^{-1}(\bI-\be_0\be_0^\Trans) & \text{$x$ interior}\\
f(x) \left(\bS_{p,x}^{-1}\bGamma_{p,x}\bS_{p,x}^{-1}+c\be_0\be_0^\Trans\right) & x=\xmin+ch\\
f(x) \left(\bS_{p,x}^{-1}\bGamma_{p,x}\bS_{p,x}^{-1}+ c\be_0\be_0^\Trans-(\be_1\be_0^\Trans + \be_0\be_1^\Trans)\right) & x=\xmax-ch.
\end{cases}
\end{align*}
\end{lem}

The scaling matrix depends on whether the evaluation point is located in the interior or boundary, which is a unique feature of our estimator. To see the intuition, consider an interior point $x$, and recall that the first element of $\hat{\bbeta}_p(x)$ is the smoothed EDF, which is $\sqrt{n}$-estimable. Therefore, the property of $\hat{F}_p(x)$ is very different from those of the estimated density and higher order derivatives.

When $x$ is either in the lower or upper boundary region, $\hat{F}_p(x)$ essentially estimates 0 or 1, respectively, hence it is super-consistent in the sense that it converges even faster than $1/\sqrt{n}$. In this case, the leading $1/\sqrt{n}$-variance vanishes, and higher order residual noise dominates, which makes $\hat{F}_p(x)$ no longer independent of the estimated density and derivatives, justifying the formula of boundary evaluation points. 

Finally we consider the second order U-statistic component. 
\begin{lem}\label{lem4: quadratic variance}
Assume Assumptions \ref{assumption: 1 dgp} and \ref{assumption: 2 kernel} hold, $h\to 0$ and $nh\to\infty$. Then
\begin{align*}
\Var[\hat{\bR}] &= \frac{2}{n^2h}f(x)F(x)(1-F(x))\bT_{p,x} + O(n^{-2}).
\end{align*}
In particular, when $x$ is in the boundary region, the above has order $O(n^{-2})$. 
\end{lem}

\subsection{Main Results}\label{subsection:main result}

In this section we provide two main results, one on asymptotic normality, and the other on standard error. 

\begin{thm}[Asymptotic Normality]\label{thm1: asy normality}
Assume Assumptions \ref{assumption: 1 dgp} and \ref{assumption: 2 kernel} hold with $\alpha_x\geq p+1$ for some integer $p\geq 0$. Further $h\to 0$, $nh^2\to \infty$ and $nh^{2p+1}=O(1)$. Then
\begin{align*}
\sqrt{nh^{2v-1}}\Big( \hat{F}_p^{(v)}(x) - F^{(v)}(x) - h^{p+1-v}\mathcal{B}_{p,v}(x) \Big) &\rightsquigarrow \mathcal{N}\Big(0,\ \mathcal{V}_{p,v}(x)\Big),\qquad 1\leq v\leq p,\\
\sqrt{\frac{n}{\mathcal{V}_{p,0}(x)}} \Big( \hat{F}_p(x) - F(x) - h^{p+1}\mathcal{B}_{p,0}(x) \Big) &\rightsquigarrow \mathcal{N}\Big(0,\ 1\Big).
\end{align*}
The constants are
\begin{align*}
\mathcal{B}_{p,v}(x) &= v! \frac{F^{(p+1)}(x)}{(p+1)!} \be_v^\Trans \bS_{p,x}^{-1} \bc_{p,x},
\end{align*}
and
\begin{align*}
\mathcal{V}_{p,v}(x) &= \begin{cases}
(v!)^2 f(x) \be_v^\Trans \bS_{p,x}^{-1}\bGamma_{p,x}\bS_{p,x}^{-1} \be_v \qquad& 1\leq v\leq p\\
F(x)(1 - F(x)) & \text{$v=0$, $x$ interior}\\
hf(x) \left(\be_0^\Trans\bS_{p,x}^{-1}\bGamma_{p,x}\bS_{p,x}^{-1}\be_0+c\right) & \text{$v=0$, $x=\xmin+ch$ or $\xmax-ch$}.
\end{cases}
\end{align*}
\end{thm}

\begin{remark}[On $nh^{2p+1}=O(1)$]
This condition ensures that higher order bias, after scaling, is asymptotically negligible. \qquad \qquad $\parallel$
\end{remark}

\begin{remark}[On $nh^2\to \infty$]
This condition ensures that the second order U-statistic, $\hat{\bR}$, has smaller order compared to $\hat{\bL}$. Note that this condition can be dropped for boundary $x$ or when the parameter of interest is the CDF $\hat{F}_p$. \qquad \qquad $\parallel$
\end{remark}

Now we provide a standard error, which is also boundary adaptive. Given the formula in Theorem \ref{thm1: asy normality}, it is possible to estimate the asymptotic variance by plugging in unknown quantities regarding the data generating process. For example consider $\mathcal{V}_{p,1}(x)$ for the estimated density. Assume the researcher knows the location of the boundary $\xmin$ and $\xmax$, the matrices $\bS_{p,x}$ and $\bGamma_{p,x}$ can be constructed with numerical integration, since they are related to features of the kernel function, not the data generating process. The unknown density $f(x)$ can also be replaced by its estimate, as long as $p\geq 1$. 

Another approach is to utilize the decomposition of the estimator, in particular the $\hat{\bL}$ term. To introduce our variance estimator, we make the following definitions. 
\begin{align*}
\hat{\bS}_{p,x} &= \frac{1}{n}\bX_h\bK_h\bX_n = \frac{1}{n}\sum_i \br_p\left( \frac{x_i-x}{h} \right)\br_p\left( \frac{x_i-x}{h} \right)^\Trans K_h(x_i-x)\\
\hat{\bGamma}_{p,x} &= \frac{1}{n^3}\sum_{i,j,k} \br_p\left( \frac{x_j-x}{h} \right)\br_p\left( \frac{x_k-x}{h} \right)^\Trans K_h(x_j-x)K_h(x_k-x)\Big( \Indicator[x_i\leq x_j]-\tilde{F}(x_j) \Big)\Big( \Indicator[x_i\leq x_k]-\tilde{F}(x_k) \Big).
\end{align*}
Following is the main result regarding variance estimation. It is automatic and fully-adaptive, in the sense that no knowledge about the boundary location is needed. 

\begin{thm}[Variance Estimation]\label{thm2: variance estimation}\ \\
Assume Assumptions \ref{assumption: 1 dgp} and \ref{assumption: 2 kernel} hold with $\alpha_x\geq p+1$ for some integer $p\geq 0$. Further $h\to 0$, $nh^2\to \infty$ and $nh^{2p+1}=O(1)$. Then
\begin{align*}
\hat{\mathcal{V}}_{p,v}(x) \equiv (v!)^2 \be_v^\Trans \bN_x \hat{\bS}_{p,x}^{-1}\hat{\bGamma}_{p,x}\hat{\bS}_{p,x}^{-1} \bN_x \be_v\ \to_\Prob\  \mathcal{V}_{p,v}(x).
\end{align*}
Define the standard error as
\begin{align*}
\hat{\sigma}_{p,v}(x) &\equiv (v!) \sqrt{\frac{1}{nh^{2v}} \be_v^\Trans  \hat{\bS}_{p,x}^{-1}\hat{\bGamma}_{p,x}\hat{\bS}_{p,x}^{-1}  \be_v}, 
\end{align*}
then
\begin{align*}
\hat{\sigma}_{p,v}(x)^{-1} \Big( \hat{F}_p^{(v)}(x) - F^{(v)}(x) - h^{p+1-v}\mathcal{B}_{p,v}(x) \Big) \rightsquigarrow \mathcal{N}\Big( 0,\ 1\Big).
\end{align*}
\end{thm}

\section{Bandwidth Selection}

In this section we consider the problem of constructing MSE-optimal bandwidth for our local polynomial regression-based distribution estimators. We focus exclusively on the case $v\geq 1$, hence the object of interest will be either the density function or derivatives thereof. Valid bandwidth choice for the distribution function $\hat{F}_p(x)$ is also an interesting topic, but difficulty arises since it is estimated at the parametric rate. We will briefly mention MSE expansion of the estimated CDF at the end. 

\subsection{For Density and Derivatives Estimates ($v\geq 1$)}

Consider some $1\leq v\leq p$, the following lemma gives finer characterization of the bias. 

\begin{lem}\label{lem5: higher order bias}
Assume Assumptions \ref{assumption: 1 dgp} and \ref{assumption: 2 kernel} hold with $\alpha_x\geq p+2$, $h\to 0$ and $nh^3\to \infty$. Then the leading bias of $\hat{F}_p^{(v)}(x)$ is
\begin{align*}
h^{p+1-v}\mathcal{B}_{p,v}(x) &= h^{p+1-v}\left\{ 
\frac{F^{(p+1)}(x)}{(p+1)!}v!\be_v^\Trans \bS_{p,x}^{-1}\bc_{p,x} + 
h\left( \frac{F^{(p+2)}(x)}{(p+2)!} + \frac{F^{(p+1)}(x)}{(p+1)!}\frac{F^{(2)}(x)}{f(x)}  \right)v!\be_v^\Trans \bS_{p,x}^{-1}\tilde{\bc}_{p,x} \right\}.
\end{align*}
\end{lem}

The above lemma is a refinement of Lemma \ref{lem1: denominator} and \ref{lem2: smoothing bias}, and characterizes the higher-order bias. To see its necessity, we note that when $p-v$ is even and $x$ is an interior evaluation point, the leading bias is zero. This is because $\be_v^\Trans \bS_{p,x}^{-1}\bc_{p,x}$ is zero, which is explained in \cite{Fan-Gijbels_1996_Book}. Except for rare cases such as $F^{(p+1)}(x)=0$ or $F^{(p+2)}(x)=0$, we have
\begin{center}
Order of bias: $h^{p+1-v}\mathcal{B}_{p,v}(x)\asymp $

\ \\
\begin{tabular}{r|cc}
\hline
& $p-v$ odd & even\\ \hline
$x$ interior & $h^{p+1-v}$ & $h^{p+2-v}$\\
boundary & $h^{p+1-v}$ & $h^{p+1-v}$\\ \hline
\end{tabular}
\end{center}
Note that for boundary evaluation points, the leading bias never vanishes. 

The leading variance is also characterized by Theorem \ref{thm1: asy normality}, and we reproduce it here:
\begin{align*}
\frac{1}{nh^{2v-1}}\mathcal{V}_{p,v}(x) &= \frac{1}{nh^{2v-1}}(v!)^2 f(x)\be_v^\Trans\bS_{p,x}^{-1}\bGamma_{p,x}\bS_{p,x}^{-1}\be_v.
\end{align*}
The MSE-optimal bandwidth is defined as a minimizer of the following
\begin{align*}
h_{ p,v}(x) &= \arg\min_{h>0} \left[\frac{1}{nh^{2v-1}}\mathcal{V}_{p,v}(x) + h^{2p+2-2v}\mathcal{B}_{p,v}(x)^2\right]. 
\end{align*}
Given the discussion we had earlier on the bias, it is easy to see that the MSE-optimal bandwidth has the following asymptotic order:
\begin{center}
Order of MSE-optimal bandwidth: $h_{ p,v}(x)\asymp $

\ \\
\begin{tabular}{r|cc}
\hline
& $p-v$ odd & even\\ \hline
$x$ interior & $n^{-\frac{1}{2p+1}}$ & $n^{-\frac{1}{2p+3}}$\\
boundary & $n^{-\frac{1}{2p+1}}$ & $n^{-\frac{1}{2p+1}}$\\ \hline
\end{tabular}
\end{center}
Again only the case where $p-v$ is even and $x$ is interior needs special attention. 

There are two notions of bandwidth consistency. Let $h$ be some non-stochastic bandwidth sequence, and $\hat{h}$ be an estimated bandwidth. Then $\hat{h}$ is consistent \emph{in rate} if $\hat{h}\asymp h$ (in most cases it is even true that $\hat{h}/h\to_\Prob C\in(0,\infty)$). And $\hat{h}$ is consistent \emph{in rate and constant} if $\hat{h}/h\to_\Prob 1$. 

To construct consistent bandwidth, either rate consistent or consistent in both rate and constant, we need estimates of both the bias and variance. The variance part is relatively easy, as we have already demonstrated in Theorem \ref{thm2: variance estimation}:
\begin{align*}
n\ell^{2v-1}\frac{\hat{\sigma}_{p,v}(x)^2}{\mathcal{V}_{p,v}(x)} \to_\Prob 1, 
\end{align*}
where $\ell$ is some preliminary bandwidth used to construct $\hat{\sigma}_{p,v}(x)$.

To introduce our bias estimate, first assume there are consistent estimators for $F^{(p+1)}(x)$ and $F^{(p+2)}(x)$, denoted by $\hat{F}^{(p+1)}(x)$ and $\hat{F}^{(p+2)}(x)$. They can be obtained, for example, using our local polynomial regression-based approach, or can be constructed with some reference model (such as the normal distribution). The critical step is to obtain consistent estimators of the matrices, which are given in the following lemma. 

\begin{lem}\label{lem6: bias estimation}
Assume Assumptions \ref{assumption: 1 dgp} and \ref{assumption: 2 kernel} hold, $\ell\to 0$ and $n\ell\to \infty$. Then
\begin{align*}
\widehat{\bS_{p,x}^{-1}\bc_{p,x}}
&= \left(\frac{1}{n}\sum_i\br_p\left(\frac{x_i-x}{\ell}\right)\br_p\left(\frac{x_i-x}{\ell}\right)^\Trans K_\ell(x_i-x) \right)^{-1}\left( \frac{1}{n}\sum_i \left(\frac{x_i-x}{\ell}\right)^{p+1}\br_p\left(\frac{x_i-x}{\ell}\right)K_\ell(x_i-x) \right)\\ 
&\qquad \qquad \qquad \to_\Prob \bS_{p,x}^{-1}\bc_{p,x},
\end{align*}
and
\begin{align*}
\widehat{\bS_{p,x}^{-1}\tilde{\bc}_{p,x}} 
&= \left(\frac{1}{n}\sum_i\br_p\left(\frac{x_i-x}{\ell}\right)\br_p\left(\frac{x_i-x}{\ell}\right)^\Trans K_\ell(x_i-x) \right)^{-1}\left( \frac{1}{n}\sum_i \left(\frac{x_i-x}{\ell}\right)^{p+2}\br_p\left(\frac{x_i-x}{\ell}\right)K_\ell(x_i-x) \right)\\ 
&\qquad \qquad \qquad \to_\Prob \bS_{p,x}^{-1}\tilde{\bc}_{p,x}.
\end{align*}
\end{lem}

Note that we used different notation, $\ell$, as it corresponds to a preliminary bandwidth. Define
\begin{align*}
h^{p+1-v}\hat{\mathcal{B}}_{p,v}(x) &= h^{p+1-v}\left\{ 
\frac{\hat{F}^{(p+1)}(x)}{(p+1)!}v!\be_v^\Trans \widehat{\bS_{p,x}^{-1}\bc_{p,x}} + 
h\frac{\hat{F}^{(p+2)}(x)}{(p+2)!}v!\be_v^\Trans \widehat{\bS_{p,x}^{-1}\tilde{\bc}_{p,x}} \right\},
\end{align*}
and assume that $\hat{\sigma}_{p,v}(x)$ is constructed using the preliminary bandwidth $\ell$. Then
\begin{align*}
\hat{h}_{ p,v}(x) &= \arg\min_{h>0} \left[\frac{\ell^{2v-1}}{h^{2v-1}}\hat{\sigma}_{p,v}(x)^2+ h^{2p+2-2v}\hat{\mathcal{B}}_{p,v}(x)^2\right]. 
\end{align*}
We make some remarks here. 

\begin{remark}[Preliminary bandwidth $\ell$]
The optimization argument $h$ enters the RHS of the previous display in three places. First it is part of the variance component, by $1/h^{2v-1}$. Second it shows as a multiplicative factor of the bias component, $h^{2p-2v+2}$. Finally within the definition of $\hat{\mathcal{B}}_{p,v}(x)$, there is another multiplicative $h$, in front of the higher order bias. 

The preliminary bandwidth $\ell$, serves a different role. It is used to estimate the variance and bias components. Of course one can use different preliminary bandwidths for $\hat{\sigma}_{p,v}(x)$, $\widehat{\bS_{p,x}^{-1}\bc_{p,x}}$ and $\widehat{\bS_{p,x}^{-1}\tilde{\bc}_{p,x}} $. 
\qquad \qquad $\parallel$
\end{remark}

\begin{remark}[Consistent bias estimator]
The bias estimator we proposed, $h^{p-v+1}\hat{\mathcal{B}}_{p,v}(x)$, is consistent in rate for the true leading bias, but not necessarily in constant. Compare $\hat{\mathcal{B}}_{p,v}(x)$ and ${\mathcal{B}}_{p,v}(x)$, it is easily seen that the term involving $F^{(p+1)}(x)F^{(2)}(x)/f(x)$ is not captured. To capture this term, we need one additional nonparametric estimator for $F^{(2)}(x)$. This is indeed feasible, and one can employ our local polynomial regression-based estimator for this purpose. 
\qquad \qquad $\parallel$
\end{remark}

\begin{thm}[Consistent bandwidth]\label{thm3: bandwidth consistency}
Let $1\leq v\leq p$. Assume the preliminary bandwidth $\ell$ is chosen such that $nh^{2v-1}\hat{\sigma}_{p,v}(x)^2/\mathcal{V}_{p,v}(x) \to_\Prob 1$, $\widehat{\bS_{p,x}^{-1}\bc_{p,x}}\to_\Prob {\bS_{p,x}^{-1}\bc_{p,x}}$, and $\widehat{\bS_{p,x}^{-1}\tilde{\bc}_{p,x}} \to_\Prob {\bS_{p,x}^{-1}\tilde{\bc}_{p,x}}$. Under the conditions of Lemma \ref{lem1: denominator} and Theorem \ref{thm2: variance estimation}:
\begin{itemize}
\item If either $x$ is in boundary regions or $p-v$ is odd, let $\hat{F}^{(p+1)}(x)$ be consistent for $F^{(p+1)}\neq 0$. Then
\begin{align*}
\frac{\hat{h}_{ p,v}(x)}{{h}_{ p,v}(x)} \to_\Prob 1.
\end{align*}
\item If $x$ is in interior and $p-v$ is even, let $\hat{F}^{(p+2)}(x)$ be consistent for $F^{(p+2)}\neq 0$. Further assume $nh^3\to 0$ and ${h}_{ p,v}(x)$ is well-defined. Then
\begin{align*}
\frac{\hat{h}_{ p,v}(x)}{{h}_{ p,v}(x)} \to_\Prob C\in(0,\infty).
\end{align*}
\end{itemize}
\end{thm}

\subsection{For CDF Estimate ($v=0$)}

In this subsection we mention briefly how to choose bandwidth for the CDF estimate, $\hat{F}_p^{(0)}(x)\equiv \hat{F}_p(x)$. We assume $x$ is in interior. Previous discussions on bias remains to apply: 
\begin{align*}
h^{p+1}\mathcal{B}_{p,0}(x) &= h^{p+1}\left\{ 
\frac{F^{(p+1)}(x)}{(p+1)!}\be_0^\Trans \bS_{p,x}^{-1}\bc_{p,x} + 
h\left( \frac{F^{(p+2)}(x)}{(p+2)!} + \frac{F^{(p+1)}(x)}{(p+1)!}\frac{F^{(2)}(x)}{f(x)}  \right)\be_0^\Trans \bS_{p,x}^{-1}\tilde{\bc}_{p,x} \right\},
\end{align*}
which means the bias of $\hat{F}_p(x)$ has order $h^{p+1}$ if either $x$ is boundary or $p$ is odd, and $h^{p+2}$ otherwise. Difficulty arises since the CDF estimator has leading variance of order
\begin{align*}
\mathcal{V}_{p,0}(x) \asymp \frac{\Indicator[\text{$x$ interior}] + h}{n},
\end{align*}
which cannot be used for bandwidth selection, because the above is proportional to the bandwidth (i.e., there is no bias-variance trade-off). 

The trick is to use a higher order variance term. Recall that the local polynomial regression-based estimator is essentially a second order U-statistic, which is then decomposed into two terms, a linear term $\hat{\bL}$ and a quadratic term $\hat{\bR}$, where the latter is a degenerate second-order U-statistic. The variance of the quadratic term $\hat{\bR}$ has been ignored so far, as it is negligible compared to the variance of the linear term. For the CDF estimator, however, it is the variance of this quadratic term that leads to a bias-variance trade-off. The exact form of this variance is given in Lemma \ref{lem4: quadratic variance}. With this additional variance term included, we have (with some abuse of notation)
\begin{align*}
\mathcal{V}_{p,0}(x) \asymp \frac{\Indicator[\text{$x$ interior}] + h}{n} + \frac{\Indicator[\text{$x$ interior}] + h}{n^2h}.
\end{align*}
Provided $x$ is an interior point, the additional variance term increases as the bandwidth shrinks. As a result, a MSE-optimal bandwidth for $\hat{F}_p(x)$ is well-defined, and estimating this bandwidth is also straightforward.
\begin{center}
Order of MSE-optimal bandwidth: $h_{ p,0}(x)\asymp $

\ \\
\begin{tabular}{r|cc}
\hline
& $p-v$ odd & even\\ \hline
$x$ interior & $n^{-\frac{2}{2p+3}}$ & $n^{-\frac{2}{2p+5}}$\\
boundary & undefined & undefined\\ \hline
\end{tabular}
\end{center}

What if $x$ is in a boundary region? Then the MSE-optimal bandwidth for $\hat{F}_p(x)$ is not well defined. The leading variance now takes the form $h/n + 1/n^2$, which is proportional to the bandwidth. (This is not surprising, since for boundary $x$ the CDF is known, and a very small bandwidth gives a super-consistent estimator.). Although MSE-optimal bandwidth for $\hat{F}_p(x)$ is not well-defined for boundary $x$, it is still feasible to minimize the empirical MSE. To see how this works, one first estimate the bias term and variance term with some preliminary bandwidth $\ell$, leading to $\hat{\mathcal{B}}_{p,0}(x)$ and $\hat{\mathcal{V}}_{p,0}(x)$. Then the MSE-optimal bandwidth can be constructed by minimizing the empirical MSE. Under regularity conditions, $\hat{\mathcal{B}}_{p,0}(x)$ will converge to some nonzero constant, while, if $x$ is boundary, $\hat{\mathcal{V}}_{p,0}(x)$ has order $\ell$, the same as the preliminary bandwidth. Then the MSE-optimal bandwidth constructed in this way will have the following order: 
\begin{center}
Order of estimated MSE-optimal bandwidth: $\hat{h}_{ p,0}(x)\asymp $

\ \\
\begin{tabular}{r|cc}
\hline
& $p-v$ odd & even\\ \hline
$x$ interior & $n^{-\frac{2}{2p+3}}$ & $n^{-\frac{2}{2p+5}}$\\
boundary & $(n^2/\ell)^{-\frac{1}{2p+3}}$ & $(n^2/\ell)^{-\frac{1}{2p+5}}$\\ \hline
\end{tabular}
\end{center}
Note that the preliminary bandwidth enters the rate of $\hat{h}_{ p,0}(x)$ for boundary $x$, because it determines the rate at which the variance estimator $\hat{\mathcal{V}}_{p,0}(x)$ vanishes. Although this estimated bandwidth is not consistent for any well-defined object, it can be useful in practice, and it reflects the fact that for boundary $x$ it is appropriate to use bandwidth shrinks fast when the object of interest is the CDF

\section{Application to Manipulation Testing}

We devote this section to density discontinuity (manipulation) tests in regression discontinuity designs. Assume there is a natural (and known) partition of the support $\mathcal{X}=[\xmin,\xmax] = [\xmin,\cutoff)\cup [\cutoff, \xmax] = \mathcal{X}_-\cup \mathcal{X}_+$, and the regularity conditions we imposed so far are satisfied on each of the partitions, $\mathcal{X}_-$ and $\mathcal{X}_+$. To be precise, assume the distribution $F$ is continuously differentiable to a certain order on each of the partitions, but the derivatives are not necessarily continuous across the cutoff $\cutoff$. In this case consistent estimates of the densities (and derivatives thereof) require fitting local polynomials separately on each sides of $\cutoff$. Alternatively, one can use the joint estimation framework introduced below. 

\subsection{Unrestricted Model}

By an unrestricted model with cutoff $\cutoff$, we consider the following polynomial basis $\br_p$
\begin{align*}
\br_p(u) = \begin{bmatrix}
\mathbf{1}_{\{u< 0\}} & u\mathbf{1}_{\{u< 0\}} & \cdots & u^p\mathbf{1}_{\{u< 0\}} & \Big| & \mathbf{1}_{\{u\geq 0\}} & u\mathbf{1}_{\{u\geq 0\}} & \cdots & u^p\mathbf{1}_{\{u\geq 0\}}
\end{bmatrix}'\in \mathbb{R}^{2p+2}. 
\end{align*}
The following two vectors will arise later, which we give the definition here:
\begin{align*}
\br_{-,p}(u) &= \begin{bmatrix}
1 & u & \cdots & u^p & 0 & \cdots & 0
\end{bmatrix}',\quad \br_{+,p}(u) = \begin{bmatrix}
0 & 0 & \cdots & 0 & \ 1 & \cdots & u^p
\end{bmatrix}'.
\end{align*}
Also we define the vectors to extract the corresponding derivatives
\begin{align*}
\bI_{2p+2} = \begin{bmatrix}
\be_{0,-} & \be_{1,-} & \cdots & \be_{p,-} & \be_{0,+} & \be_{1,+} & \cdots & \be_{p,+}
\end{bmatrix}.
\end{align*}
With the above definition, the estimator at the cutoff is\footnote{The EDF is defined with the whole sample as before: $\tilde{F}(u)=n^{-1}\sum_i \Indicator[x_i\leq u]$.}
\begin{align*}
\hat \bbeta_p(\cutoff) = \arg\min_{\mathbf{b}\in\mathbb{R}^{2p+2}}\sum_{i} \Big( \tilde{F}(x_i) -\br_p(x_i-\cutoff)'\mathbf{b}  \Big)^2 K_{h}(x_i-\cutoff).
\end{align*}
We assume the same bandwidth is used below and above the cutoff to avoid cumbersome notation. Generalizing to using different bandwidths is straightforward. Other notations (for example $\bX$ and $\bX_h$) are redefined similarly, with the scaling matrix $\bH$ adjusted so that $\bH^{-1}\br_p(u) = \br_p(h^{-1}u)$ is always true. we denote the estimates by
\begin{align*}
\hat{F}^{(v)}_p(\cutoff-) &= v!\be_{v,-}'\hat{\bbeta}_p(\cutoff),\qquad \hat{F}^{(v)}_p(\cutoff+) = v!\be_{v,+}'\hat{\bbeta}_p(\cutoff).
\end{align*}

\begin{remark}[Separate estimation]
An alternative implementation is to apply our local polynomial-based estimator separately to the two samples, one with observations below the cutoff, and the other with observations above the cutoff. To be precise, let $\tilde{F}_{-}(\cdot)$ and $\tilde{F}_{+}(\cdot)$ be the empirical distribution functions constructed by the two samples. That is, 
\begin{align*}
\tilde{F}_{-}(x) &= \frac{1}{n_-}\sum_{i:\ x_i<\cutoff}\Indicator[x_i\leq x],\qquad \tilde{F}_{+}(x) = \frac{1}{n_+}\sum_{i:\ x_i\geq\cutoff}\Indicator[x_i\leq x],
\end{align*}  
where $n_-$ and $n_+$ denote the size of the two samples, respectively. The the local polynomial approach, applied to $\tilde{F}_{-}(\cdot)$ and $\tilde{F}_{+}(\cdot)$ separately, will yield two sets of estimates, which we denote by $\hat{F}_{p,-}^{(v)}(\cutoff)$ and $\hat{F}_{p,+}^{(v)}(\cutoff)$. To see the relation between joint and separate estimations, we note the following (which can be easily seen using least squares algebra)
\begin{alignat*}{3}
v=0     &\qquad \hat{F}_{p,-}(\cutoff) &&= \frac{n}{n_-}\hat{F}_{p}(\cutoff-),\quad &&\hat{F}_{p,+}(\cutoff) = \frac{n}{n_+}\hat{F}_{p}(\cutoff+) - \frac{n_-}{n_+}\\
v\geq 1 &\qquad \hat{F}_{p,-}^{(v)}(\cutoff) &&= \frac{n}{n_-}\hat{F}_{p}^{(v)}(\cutoff-),\quad &&\hat{F}_{p,+}^{(v)}(\cutoff) = \frac{n}{n_+}\hat{F}_{p}^{(v)}(\cutoff+).
\end{alignat*}
The difference comes from the fact that by separate estimation, one obtains estimates of the conditional CDF and the derivatives. 
\qquad \qquad $\parallel$
\end{remark}

In the following lemmas, we will give asymptotic results for the joint estimation problem. Proofs are omitted.

\begin{lem}\label{lem7: denominator, unrestricted}
Let Assumptions of Lemma \ref{lem1: denominator} hold separately on $\mathcal{X}_-$ and $\mathcal{X}_+$, then
\begin{align*}
\nonumber\frac{1}{n}\bX_h'\bK_h\bX_h &= f(\cutoff-)\bS_{-,p} +f(\cutoff+)\bS_{+,p}  + O\Big(h\Big) + \Op\Big(1/\sqrt{nh}\Big),
\end{align*}
where 
\begin{align*}
\bS_{-,p} = \int_{-1}^0 \br_{-,p}(u)\br_{-,p}(u)'K(u) \diff u,\qquad \bS_{+,p} = \int_0^1 \br_{+,p}(u)\br_{+,p}(u)'K(u) \diff u.
\end{align*}
\end{lem} 

Again we decompose the estimator into four terms, namely $\hat{\bB}_{\mathtt{LI}}$, $\hat{\bB}_{\mathtt{S}}$, $\hat{\bL}$ and $\hat{\bR}$. 

\begin{lem}\label{lem8: smoothing bias, unrestricted}
Let Assumptions of Lemma \ref{lem2: smoothing bias} hold separately on $\mathcal{X}_-$ and $\mathcal{X}_+$, then
\begin{align*}
\hat{\bB}_{\mathtt{S}}&= h^{p+1} \left\{ \frac{F^{(p+1)}(\cutoff-)f(\cutoff-)}{(p+1)!}\bc_{-,p} +  \frac{F^{(p+1)}(\cutoff+)f(x+)}{(p+1)!}\bc_{+,p}  \right\}+ \op(h^{p+1}),\qquad \hat{\bB}_{\mathtt{LI}} = \Op\left(\frac{1}{n}\right),
\end{align*}
where
\begin{align*}
\bc_{-,p} = \int_{-1}^{0} u^{p+1}\br_{-,p}(u)K(u)\diff u,\qquad \bc_{+,p} = \int_{0}^{1} u^{p+1}\br_{+,p}(u)K(u)\diff u.
\end{align*}
\end{lem}

\begin{lem}\label{lem9: linear variance, unrestricted}
Let Assumptions of Lemma \ref{lem3: linear variance} hold separately on $\mathcal{X}_-$ and $\mathcal{X}_+$, then
\begin{align*}
\Var\left[ \sqrt{\frac{n}{h}}\Big(\be_{1,+}-\be_{1,-}\Big)'\Big(f(\cutoff+)\bS_{+,p} +f(\cutoff-)\bS_{-,p}\Big)^{-1}\hat{\bL} \right]
&= f(\cutoff-) \be_{1,-}'\bS_{-,p}^{-1}\bGamma_{-,p}\bS_{-,p}^{-1}\be_{1,-} \\
&\qquad+ f(\cutoff+) \be_{1,+}'\bS_{+,p}^{-1}\bGamma_{+,p}\bS_{+,p}^{-1}\be_{1,+} + O(h),
\end{align*}
where 
\begin{align*}
\bGamma_{-,p} &= \iint_{[-1,0]^2} (u\wedge v)\br_{-,p}(u)\br_{-,p}(v)'K(u)K(v)\ \diff u\diff v,\quad \bGamma_{+,p} = \iint_{[0,1]^2} (u\wedge v)\br_{+,p}(u)\br_{+,p}(v)'K(u)K(v)\ \diff u\diff v.
\end{align*}
\end{lem}

Note that the above gives the asymptotic variance of the difference $\hat{f}(\cutoff+) - \hat{f}(\cutoff-)$, and the variance takes an additive form. This is not surprising, since the two density estimates, $\hat{f}(\cutoff+)$ and $\hat{f}(\cutoff-)$, rely on distinctive subsamples, meaning that they are asymptotically independent. 

Finally the order of $\hat{\bR}$ can also be established. 

\begin{lem}\label{lem10: quadratic variance, unrestricted}
Let Assumptions of Lemma \ref{lem4: quadratic variance} hold separately on $\mathcal{X}_-$ and $\mathcal{X}_+$, then 
\begin{align*}
\hat{\bR} = \Op\left( \sqrt{\frac{1}{n^2h}} \right).
\end{align*}
\end{lem}

Now we state the main result concerning the manipulation testing. Let $\hat{\bS}_{p,\cutoff}$ and $\hat{\bGamma}_{p,\cutoff}$ be constructed as in Section \ref{subsection:main result}, and 
\begin{align*}
\hat{\mathcal{V}}_{p,1}(\cutoff) &= \frac{1}{h}(\be_{1,+} - \be_{1,-})'\hat{\bS}_{p,\cutoff}\hat{\bGamma}_{p,\cutoff} \hat{\bS}_{p,\cutoff}(\be_{1,+} - \be_{1,-}).
\end{align*}  

\begin{coro}\label{coro1: asy normal, unrestricted}
Assume Assumptions \ref{assumption: 1 dgp} and \ref{assumption: 2 kernel} hold separately on $\mathcal{X}_-$ and $\mathcal{X}_+$ with $\alpha_x\geq p+1$ for some integer $p\geq 1$. Further, $n \cdot h^2\to \infty$ and $n\cdot h^{2p+1}\to 0$. Then under the null hypothesis $\mathsf{H}_0:f(\cutoff+) = f(\cutoff-)$,
\begin{align*}
T_p(h) &= \frac{\hat{f}_p(\cutoff+) - \hat{f}_p(\cutoff-) }{\sqrt{\frac{1}{nh}\hat{\mathcal{V}}_{p,1}(\cutoff)}} \rightsquigarrow \mathcal{N}(0,1). 
\end{align*}
As a result, under the alternative hypothesis $\mathsf{H}_1:f(\cutoff+) \neq f(\cutoff-)$,
\begin{align*}
\lim_{n\to\infty}\Prob[|T_p(h)| \geq \Phi_{1-\alpha/2}] = 1.
\end{align*}
Here $\Phi_{1-\alpha/2}$ is the $(1-\alpha/2)$-quantile of the standard normal distribution. 
\end{coro}

\begin{remark}[Separate estimation]
Recall that it is possible to implement our local polynomial estimator separately for the two subsamples, below and above the cutoff $\cutoff$. Let $\hat{f}_{p,-}(\cutoff)$ and $\hat{f}_{p,+}(\cutoff)$ be the two density estimates, and $\hat{\mathcal{V}}_{p,1,-}(\cutoff)$ and $\hat{\mathcal{V}}_{p,1,+}(\cutoff)$ be the associated variance estimates. Then the test statistic is equivalently:
\begin{align*}
T_p(h) &= \frac{\frac{n_+}{n}\hat{f}_{p,+}(\cutoff) - \frac{n_-}{n}\hat{f}_{p,-}(\cutoff) }{\sqrt{\frac{1}{nh}\left(\frac{n_+}{n}\hat{\mathcal{V}}_{p,1,+}(\cutoff)+\frac{n_-}{n}\hat{\mathcal{V}}_{p,1,-}(\cutoff)\right)}}.
\end{align*}
$\parallel$
\end{remark}

\subsection{Restricted Model}

In the previous subsection, we gave a test procedure on the discontinuity of the density by estimating on the two sides of the cutoff separately. This procedure is flexible and requires minimum assumptions. There are ways, however, to improve the power of the test when the densities are estimated with additional assumptions on the smoothness of the CDF 

In a restricted model, the polynomial basis is re-defined as 
\begin{align*}
\br_p(u) = \begin{bmatrix}
1 & u\mathbf{1}(u< 0) & u\mathbf{1}(u\geq 0) & u^2 & u^3 & \cdots & u^p
\end{bmatrix}'\in\mathbb{R}^{p+2},
\end{align*}
and the estimator in the fully restricted model is
\begin{align*}
\hat{\bbeta}_p(\cutoff) &=\begin{bmatrix}
\hat F_p(\cutoff) & \hat f_{p}(\cutoff-) & \hat f_{p}(\cutoff+) & \frac{1}{2}\hat F_p^{(2)}(\cutoff) & \cdots & \frac{1}{p!}\hat F_p^{(p)}(\cutoff)
\end{bmatrix}'=\arg\max_{\mathbf{b}\in\mathbb{R}^{p+2}} \sum_{i} \Big( \tilde{F}(x_i)-\br_p(x_i-\cutoff)'\mathbf{b} \Big)^2K_{h}(x_i-\cutoff).
\end{align*}
Again the notations (for example $\bX$ and $\bX_h$) are redefined similarly, with the scaling matrix $\bH$ adjusted to ensure $\bH^{-1}\br_p(u) = \br_p(h^{-1}u)$. Here $\hat F_p(\cutoff)$ is the estimated CDF and $\frac{1}{2}\hat F_p^{(2)}(\cutoff)$, $\cdots$, $\frac{1}{p!}\hat F_p^{(p)}(\cutoff)$ are the estimated higher order derivatives, which we assume are all continuous at $\cutoff$, while $\hat f_{p}(\cutoff-)$ and $\hat f_{p}(\cutoff+)$ are the estimated densities on the two sides of $\cutoff$. Therefore we call the above model restricted, since it only allows discontinuity of the first derivative of $F$ (i.e. the density) but not the other derivatives. 

With the modification of the polynomial basis, all other matrices in the previous subsection are redefined similarly, and
\begin{align*}
\bI_{p+2} = \begin{bmatrix}
\be_0 & \be_{1,-} & \be_{1,+} & \be_2 & \cdots & \be_p
\end{bmatrix}_{(p+2)\times (p+2)}.
\end{align*}
where the subscripts indicate the corresponding derivatives to extract. Moreover 
\begin{align*}
\br_{-,p}(u) = \begin{bmatrix}
1 & u & 0 & u^2 & \cdots & u^p
\end{bmatrix},\qquad \br_{+,p}(u) = \begin{bmatrix}
1 & 0 & u & u^2 & \cdots & u^p
\end{bmatrix}.
\end{align*}

\begin{lem}\label{lem11: denominator restricted}
Let Assumptions of Lemma \ref{lem1: denominator} hold with the exception that $f$ may be discontinuous across $\cutoff$, then
\begin{align*}
 \nonumber\frac{1}{n}\bX_h'\bK_h\bX_h &= \left\{f(\cutoff-)\bS_{-,p} +f(\cutoff+)\bS_{+,p}  \right\} + O\left(h\right) + \Op(1/\sqrt{nh}),
\end{align*}
where 
\begin{align*}
\bS_{-,p} = \int_{-1}^0 \br_{-,p}(u)\br_{-,p}(u)'K(u) \diff u,\qquad \bS_{+,p} = \int_0^1 \br_{+,p}(u)\br_{+,p}(u)'K(u) \diff u.
\end{align*}
\end{lem} 

Again we decompose the estimator into four terms, $\hat{\bB}_{\mathtt{LI}}$, $\hat{\bB}_{\mathtt{S}}$, $\hat{\bL}$ and $\hat{\bR}$, which correspond to leave-in bias, smoothing bias, linear variance and quadratic variance, respectively.

\begin{lem}\label{lem12: smoothing bias, restricted}
Let Assumptions of Lemma \ref{lem2: smoothing bias} hold with the exception that $f$ may be discontinuous across $\cutoff$, then
\begin{align}
\hat{\bB}_{\mathtt{S}}&= h^{p+1} \left\{ \frac{F^{(p+1)}(\cutoff-)f(\cutoff-)}{(p+1)!}\bc_{-,p} +  \frac{F^{(p+1)}(\cutoff+)f(\cutoff+)}{(p+1)!}\bc_{+,p}  \right\}+\op(h^{p+1}),\qquad \hat{\bB}_{\mathtt{LI}}=\Op\left(\frac{1}{n}\right),
\end{align}
where
\begin{align*}
\bc_{-,p} = \int_{-1}^{0} u^{p+1}\br_{-,p}(u)K(u)\diff u,\qquad \bc_{+,p} = \int_{0}^{1} u^{p+1}\br_{+,p}(u)K(u)\diff u.
\end{align*}
\end{lem}

\begin{lem}\label{lem13: linear variance, restricted}
Let Assumptions of Lemma \ref{lem3: linear variance} hold with the exception that $f$ may be discontinuous across $\cutoff$, then
\begin{align*}
&\Var\left[ \sqrt{\frac{n}{h}}\Big(\be_{1,+}-\be_{1,-}\Big)'\Big(f(\cutoff+)\bS_{+,p} +f(\cutoff-)\bS_{-,p}\Big)^{-1}\hat{\bL} \right]\\
&\qquad = (\be_{1,+}-\be_{1,-})'(f(\cutoff+)\bS_{+,p}+f(\cutoff-)\bS_{-,p})^{-1} 
(f(\cutoff+)^3 \bGamma_{+,p}\\
&\qquad\qquad+  f(\cutoff-)^3 \boldsymbol{\Psi}\bGamma_{+,p}\boldsymbol{\Psi})
(f(\cutoff+)\bS_{+,p}+f(\cutoff-)\bS_{-,p})^{-1}(\be_{1,+}-\be_{1,-}) + O(h),
\end{align*}
where 
\begin{align*}
\bGamma_{-,p} &= \iint_{[-1,0]^2} (u\wedge v)\br_{-,p}(u)\br_{-,p}(v)'K(u)K(v)\ \diff u\diff v,\quad \bGamma_{+,p} = \iint_{[0,1]^2} (u\wedge v)\br_{+,p}(u)\br_{+,p}(v)'K(u)K(v)\ \diff u\diff v.
\end{align*}
and 
\begin{align*}
\boldsymbol{\Psi} = \begin{bmatrix}
(-1)^0 &   &   &  &  &  & \\
  &   & (-1)^1&  &  &  & \\
  & (-1)^1&   &  &  &  & \\
  &   &   & (-1)^2&  &  & \\
  &   &   &  &(-1)^3&  & \\
  &   &   &  &  & \ddots & \\
  &   &   &  &  &  & (-1)^p 
\end{bmatrix}.
\end{align*}
\end{lem}

Again we can show that the quadratic part is negligible. 

\begin{lem}\label{lem14: quadratic variance, restricted}
Let Assumptions of Lemma \ref{lem4: quadratic variance} hold with the exception that $f$ may not be continuous across $\cutoff$, then
\begin{align*}
\hat{\bR} = \Op\left( \sqrt{\frac{1}{n^2h}} \right).
\end{align*}
\end{lem}

Now we state the main result concerning the manipulation testing. Let $\hat{\bS}_{p,\cutoff}$ and $\hat{\bGamma}_{p,\cutoff}$ be constructed as in Section \ref{subsection:main result}, and 
\begin{align*}
\hat{\mathcal{V}}_{p,1}(\cutoff) &= \frac{1}{h}(\be_{1,+} - \be_{1,-})'\hat{\bS}_{p,\cutoff}\hat{\bGamma}_{p,\cutoff} \hat{\bS}_{p,\cutoff}(\be_{1,+} - \be_{1,-}).
\end{align*}  

\begin{coro}\label{coro2: asy normal, restricted}
Assume Assumptions \ref{assumption: 1 dgp} and \ref{assumption: 2 kernel} hold separately on $\mathcal{X}_-$ and $\mathcal{X}_+$ with $\alpha_x\geq p+1$ for some integer $p\geq 1$. Further, $n \cdot h^2\to \infty$ and $n\cdot h^{2p+1}\to 0$. Then under the null hypothesis $\mathsf{H}_0:f(\cutoff+) = f(\cutoff-)$,
\begin{align*}
T_p(h) &= \frac{\hat{f}_p(\cutoff+) - \hat{f}_p(\cutoff-) }{\sqrt{\frac{1}{nh}\hat{\mathcal{V}}_{p,1}(\cutoff)}} \rightsquigarrow \mathcal{N}(0,1). 
\end{align*}
As a result, under the alternative hypothesis $\mathsf{H}_1:f(\cutoff+) \neq f(\cutoff-)$,
\begin{align*}
\lim_{n\to\infty}\Prob[|T_p(h)| \geq \Phi_{1-\alpha/2}] = 1.
\end{align*}
Here $\Phi_{1-\alpha/2}$ is the $(1-\alpha/2)$-quantile of the standard normal distribution. 
\end{coro}

\section{Other Standard Error Estimators}

The standard error $\hat{\sigma}_{p,v}(x)$ (see Theorem \ref{thm2: variance estimation}) is fully automatic and adapts to both interior and boundary regions. In this section we consider two other ways to construct a standard error. 

\subsection{Plug-in Standard Error}

Take $v\geq 1$. Then the asymptotic variance of $\hat{F}^{(v)}_p(x)$ takes the following form:
\begin{align*}
\mathcal{V}_{p,v}(x) &= 
(v!)^2 f(x) \be_v^\Trans \bS_{p,x}^{-1}\bGamma_{p,x}\bS_{p,x}^{-1} \be_v. 
\end{align*}
One way of constructing estimate of the above quantity is to plug-in a consistent estimator of $f(x)$, which is simply the estimated density. Hence we can use
\begin{align*}
\hat{\mathcal{V}}_{p,v}(x) &= 
(v!)^2 \hat{f}_p(x) \be_v^\Trans \bS_{p,x}^{-1}\bGamma_{p,x}\bS_{p,x}^{-1} \be_v. 
\end{align*}
The next question is how $\bS_{p,x}$ and $\bGamma_{p,x}$ should be constructed. Note that they are related to the kernel, evaluation point $x$ and the bandwidth $h$, but \textit{not} the data generating process. Therefore the three matrices can be constructed by either analytical integration or numerical method. 

\subsection{Jackknife-based Standard Error}

The standard error $\hat{\sigma}_{p,v}(x)$ is obtained by inspecting the asymptotic linear representation. It is fully automatic and adapts to both interior and boundaries. In this part, we present another standard error which resembles $\hat{\sigma}_{p,v}(x)$, albeit with a different motivation. 

Recall that $\hat{\bbeta}_p(x)$ is essentially a second order U-statistic, and the following expansion is justified:
\begin{align*}
&\frac{1}{n}\bX_h'\bK_h\left(\bY-\bX\bbeta_p(x)\right)\\
&\qquad= \frac{1}{n} \sum_{i} \br_p\left(\frac{x_i-x}{h}\right)\left(\tilde{F}(x_i)-\br_p(x_i-x)'\bbeta_p(x)\right)K_{h}(x_i-x)\\
&\qquad= \frac{1}{n} \sum_{i} \br_p\left(\frac{x_i-x}{h}\right)\left(\frac{1}{n-1}\sum_{j;j\neq i}\Big(\Indicator(x_j\leq x_i)-\br_p(x_i-x)'\bbeta_p(x)\Big)\right)K_{h}(x_i-x) + \Op\left(\frac{1}{n}\right)\\
&\qquad= \frac{1}{n(n-1)} \sum_{i,j;i\neq j}\br_p\left(\frac{x_i-x}{h}\right)\Big(\Indicator(x_j\leq x_i)-\br_p(x_i-x)'\bbeta_p(x)\Big)K_{h}(x_i-x) + \Op\left(\frac{1}{n}\right),
\end{align*}
where the remainder represents leave-in bias. Note that the above could be written as a U-statistic, and to apply the Hoeffding decomposition, define
\begin{align*}
\bU(x_i,x_j)
&= \br_p\left(\frac{x_i-x}{h}\right)\Big(\Indicator(x_j\leq x_i)-\br_p(x_i-x)'\bbeta_p(x)\Big)K_{h}(x_i-x)\\ &\qquad+\br_p\left(\frac{x_j-x}{h}\right)\Big(\Indicator(x_i\leq x_j)-\br_p(x_j-x)'\bbeta_p(x)\Big)K_{h}(x_j-x),
\end{align*}
which is symmetric in its two arguments. Then 
\begin{align*}
\frac{1}{n}\bX_h'\bK_h\left(\bY-\bX\bbeta_p(x)\right) &= \Expectation\left[\bU(x_i,x_j)\right]+ \frac{1}{n} \sum_{i} \Bigg(\bU_{1}(x_i) - \Expectation\left[\bU(x_i,x_j)\right]\Bigg)\\
&+ \binom{n}{2}^{-1} \sum_{i,j;i<j} \Bigg(\bU(x_i,x_j) -  \bU_{1}(x_i) - \bU_{1}(x_j) + \Expectation\left[\bU(x_i,x_j)\right]\Bigg).
\end{align*}
Here $\bU_{1}(x_i) = \Expectation\left[\left. \bU(x_i,x_j)\right| x_i\right]$. The second line in the above display is the analogue of $\hat{\bL}$, which contributes to the leading variance, and the third line is negligible. The new standard error, we call the jackknife-based standard error, is given by the following:
\begin{align*}
\hat{\sigma}^{(\mathtt{JK})}_{p,v}(x) &\equiv (v!) \sqrt{\frac{1}{nh^{2v}} \be_v^\Trans  \hat{\bS}_{p,x}^{-1}\hat{\bGamma}^{\mathtt{JK}}_{p,x}\hat{\bS}_{p,x}^{-1}  \be_v}, 
\end{align*}
with 
\begin{align*}
\hat{\bGamma}^{\mathtt{JK}}_{p,x} &= \frac{1}{n}\sum_{i} \left(\frac{1}{n-1}\sum_{j;j\neq i} \hat\bU(x_i, x_j)\right)\left(\frac{1}{n-1}\sum_{j;j\neq i} \hat\bU(x_i, x_j)\right)' - \left( \binom{n}{2}^{-1} \sum_{i,j;i\neq j} \hat\bU(x_i, x_j) \right) \left( \binom{n}{2}^{-1} \sum_{i,j;i\neq j} \hat\bU(x_i, x_j) \right)',
\end{align*}
and
\begin{align*}
\hat\bU(x_i,x_j)
&= \br_p\left(\frac{x_i-x}{h}\right)\Big(\Indicator(x_j\leq x_i)-\br_p(x_i-x)'\hat{\bbeta}_p(x)\Big)K_{h}(x_i-x)\\
&\qquad +\br_p\left(\frac{x_j-x}{h}\right)\Big(\Indicator(x_i\leq x_j)-\br_p(x_j-x)'\hat{\bbeta}_p(x)\Big)K_{h}(x_j-x).
\end{align*}
The name jackknife comes from the fact that we use leave-one-out ``estimator'' for $\bU_{1}(x_i)$: with $x_i$ fixed, 
\begin{align*}
\text{``}\frac{1}{n-1}\sum_{j;j\neq i} \hat{\bU}(x_i,x_j) \to_\Prob \bU_{1}(x_i)\text{''}.
\end{align*}
Under the same conditions specified in Theorem \ref{thm2: variance estimation}, one can show that the jackknife-based standard error is consistent. 

\section{Simulation Study}

\subsection{DGP 1: Truncated Normal Distribution}

In this subsection, we conduct simulation study based on truncated normal distribution. To be more specific, the underlying distribution of $x_i$ is the standard normal distribution truncated below at $-0.8$. Therefore,
\begin{align*}
G(x) = F(x) &= \frac{\Phi(x)-\Phi(-0.8)}{1-\Phi(-0.8)},\qquad x\geq -0.8,
\end{align*}
and zero otherwise. Equivalently, $x_i$ has Lebesgue density $\Phi^{(1)}(x)/(1-\Phi(-0.8))$ on $[-0.8,\infty]$. 

In this simulation study, the target parameter is the density function evaluated at various points. Note that both the variance and the bias of our estimator depend on the evaluation point, and in particular, the magnitude of the bias depends on higher order derivatives of the distribution function. 

\begin{enumerate}
\item \textit{Evaluation point}. We estimate the density at $x\in\{ -0.8,\ -0.5,\ 0.5,\ 1.5 \}$. Note that $-0.8$ is the boundary point, where classical density estimators such as the kernel density estimator has high bias. The point $-0.5$, given our bandwidth choice, is fairly close to the boundary, hence should be understood as in the lower boundary region. The two points $0.5$ and $1.5$ are interior, but the curvature of the normal density is quite different at those two points, and we expect to see the estimators having different bias behaviors. 

\item \textit{Polynomial order}. We consider $p\in\{2,3\}$. For density estimation using our estimators, $p=2$ should be the default choice, since it corresponds to estimating conditional mean with local linear regression. Such choice is also recommended by \cite{Fan-Gijbels_1996_Book}, according to which one should always choose $p-s=2-1=1$ to be an odd number. We include $p=3$ for completeness. 

\item \textit{Kernel function}. For local polynomial regression, the choice of kernel function is usually not very important. We use the triangular kernel $k(u)=(1-|u|)\vee 0$. 

\item \textit{Sample size}. The sample size used consists of $n\in\{1000,\ 2000\}$. For most empirical studies employing nonparametric density estimation, the sample size is well above 1000, hence $n=2000$ is more representative. 
\end{enumerate}

Overall, we have $4\times 2\times 2 = 16$ designs, and for each design, we conduct 5000 Monte Carlo repetitions. 

We consider a grid of bandwidth choices, which correspond to multiples of the MSE-optimal bandwidth, ranging from $0.1h_{\mathtt{MSE}}$ to $2h_{\mathtt{MSE}}$. We also consider the estimated bandwidth. The MSE-optimal bandwidth, $h_{\mathtt{MSE}}$, is chosen by minimizing the asymptotic mean squared error, using the true underlying distribution.

For each design, we report the empirical bias of the estimator, $\Expectation[\hat{f}_p(x)-f(x)]$, under \underline{bias}. And empirical standard deviations, $\Var^{1/2}[\hat{f}_p(x)]$, and empirical root-MSE, under \underline{sd} and \underline{$\sqrt{\text{mse}}$}, respectively. For the standard errors constructed from the variance estimators, we report their empirical average under \underline{mean}, which should be compared to \underline{sd}. We also report the empirical rejection rate of t-statistics at 5\% nominal level, under \underline{size}. The t-statistic is $(\hat{f}_p(x)-\Expectation \hat{f}_p(x))/\text{se}$, which is exactly centered, hence rejection rate thereof is a measure of accuracy of normal approximation. 

\subsection{DGP 2: Exponential Distribution}

In this subsection, we conduct simulation study based on exponential distribution. To be more specific, the underlying distribution of $x_i$ is $F(x)=1-e^{-x}$. Equivalently, $x_i$ has Lebesgue density $e^{-x}$ for $x\geq 0$. 

In this simulation study, the target parameter is the density function evaluated at various points. Note that both the variance and the bias of our estimator depend on the evaluation point, and in particular, the magnitude of the bias depends on higher order derivatives of the distribution function. 

\begin{enumerate}
\item \textit{Evaluation point}. We estimate the density at $x\in\{ 0,\ 1,\ 1.5 \}$. Note that $0$ is the boundary point, where classical density estimators such as the kernel density estimator has high bias. The two points $1$ and $1.5$ are interior. 

\item \textit{Polynomial order}. We consider $p\in\{2,3\}$. For density estimation using our estimators, $p=2$ should be the default choice, since it corresponds to estimating conditional mean with local linear regression. Such choice is also recommended by \cite{Fan-Gijbels_1996_Book}, according to which one should always choose $p-s=2-1=1$ to be an odd number. We include $p=3$ for completeness. 

\item \textit{Kernel function}. For local polynomial regression, the choice of kernel function is usually not very important. We use the triangular kernel $k(u)=(1-|u|)\vee 0$. 

\item \textit{Sample size}. The sample size used consists of $n\in\{1000,\ 2000\}$. For most empirical studies employing nonparametric density estimation, the sample size is well above 1000, hence $n=2000$ is more representative. 
\end{enumerate}

Overall, we have $3\times 2\times 2 = 12$ designs, and for each design, we conduct 5000 Monte Carlo repetitions. 

We consider a grid of bandwidth choices, which correspond to multiples of the MSE-optimal bandwidth, ranging from $0.1h_{\mathtt{MSE}}$ to $2h_{\mathtt{MSE}}$. We also consider the estimated bandwidth. The MSE-optimal bandwidth, $h_{\mathtt{MSE}}$, is chosen by minimizing the asymptotic mean squared error, using the true underlying distribution.

For each design, we report the empirical bias of the estimator, $\Expectation[\hat{f}_p(x)-f(x)]$, under \underline{bias}. And empirical standard deviations, $\Var^{1/2}[\hat{f}_p(x)]$, and empirical root-MSE, under \underline{sd} and \underline{$\sqrt{\text{mse}}$}, respectively. For the standard errors constructed from the variance estimators, we report their empirical average under \underline{mean}, which should be compared to \underline{sd}. We also report the empirical rejection rate of t-statistics at 5\% nominal level, under \underline{size}. The t-statistic is $(\hat{f}_p(x)-\Expectation \hat{f}_p(x))/\text{se}$, which is exactly centered, hence rejection rate thereof is a measure of accuracy of normal approximation.

\addcontentsline{toc}{section}{References}
\bibliographystyle{jasa}
\bibliography{Cattaneo-Jansson-Ma_2019_JASA}

\clearpage

\section{Proof}
{ 

\singlespacing
\footnotesize

\subsection{Proof of Lemma \ref{lem1: denominator}}

\begin{proof}
A generic element of the matrix $\frac{1}{n}\bX_h'\bK_h\bX_h$ takes the form:
\begin{align*}
\frac{1}{n}\sum_i \frac{1}{h}\left(\frac{x_i-x}{h}\right)^s K\left(\frac{x_i-x}{h}\right),\qquad 0\leq s\leq 2p.
\end{align*}
Then we compute the expectation:
\begin{align*}
&\ \Expectation\left[ \frac{1}{n}\sum_i \frac{1}{h}\left(\frac{x_i-x}{h}\right)^s K\left(\frac{x_i-x}{h}\right) \right] = \Expectation\left[ \frac{1}{h}\left(\frac{x_i-x}{h}\right)^s K\left(\frac{x_i-x}{h}\right) \right]\\
&= \int_{\xmin}^{\xmax} \frac{1}{h}\left(\frac{u-x}{h}\right)^s K\left(\frac{u-x}{h}\right)f(u)\diff u= \int_{\xminh}^{\xmaxh} v^s K\left(v\right)f(x+vh) \diff v
= \int_{\xminh}^{\xmaxh} v^s K\left(v\right)f(x+vh) \diff v,
\end{align*}
hence for $x$ in the interior,
\begin{align*}
&\ \Expectation\left[ \frac{1}{n}\sum_i \frac{1}{h}\left(\frac{x_i-x}{h}\right)^s K\left(\frac{x_i-x}{h}\right) \right]= f(x)\int_{\mathbb{R}} \br_p(v)\br_p(v)^\Trans K(v)\diff v + o(1),
\end{align*}
and for $x=\xmin + ch$ with $c\in[0, 1]$, 
\begin{align*}
&\ \Expectation\left[ \frac{1}{n}\sum_i \frac{1}{h}\left(\frac{x_i-x}{h}\right)^s K\left(\frac{x_i-x}{h}\right) \right]= f(\xmin)\int_{-c}^\infty \br_p(v)\br_p(v)^\Trans K(v)\diff v + o(1),
\end{align*}
and for $x=\xmax - ch$ with $c\in[0, 1]$, 
\begin{align*}
&\ \Expectation\left[ \frac{1}{n}\sum_i \frac{1}{h}\left(\frac{x_i-x}{h}\right)^s K\left(\frac{x_i-x}{h}\right) \right]= f(\xmax)\int_{-\infty}^c \br_p(v)\br_p(v)^\Trans K(v)\diff v + o(1),
\end{align*}
provided that $F\in\mathcal{C}^1$. 

The variance satisfies
\begin{align*}
&\ \Var\left[ \frac{1}{n}\sum_i \frac{1}{h}\left(\frac{x_i-x}{h}\right)^s K\left(\frac{x_i-x}{h}\right) \right] = \frac{1}{n}\Var\left[\frac{1}{h}\left(\frac{x_i-x}{h}\right)^s K\left(\frac{x_i-x}{h}\right)\right]\\
&\leq \frac{1}{n}\Expectation\left[ \frac{1}{h^2}\left(\frac{x_i-x}{h}\right)^{2s} K\left(\frac{x_i-x}{h}\right)^2 \right]
= O\left(\frac{1}{nh}\right),
\end{align*}
provided that $F\in\mathcal{C}^1$. 
\end{proof}

\subsection{Proof of Lemma \ref{lem2: smoothing bias}}

\begin{proof}
First consider the smoothing bias. The leading term can be easily obtain by taking expectation together with Taylor expansion of $F$ to power $p+1$. The variance of this term has order $n^{-1}h^{-1}h^{2p+2}$, which gives the residual estimate $\op(h^{p+1})$ since it is assumed that $nh\to \infty$. 

Next for the leave-in bias, note that it has expectation of order $n^{-1}$, and variance of order $n^{-3}h^{-1}$, hence overall this term of order $\Op(n^{-1})$. 
\end{proof}

\subsection{Proof of Lemma \ref{lem3: linear variance}}

\begin{proof}
We first compute the variance. Note that 
\begin{align*}
& \int_{\xminh}^{\xmaxh}  \br_p\left(u\right)\Big(\tilde{F}(x+hu) - F(x+hu) \Big)K(u)f(x+hu)\diff u\\
&\qquad = \frac{1}{n}\int_{\xminh}^{\xmaxh}  \br_p\left(u\right) \Big(\Indicator[x_i\leq x+hu] - F(x+hu) \Big)K(u)f(x+hu)\diff u,
\end{align*}
and
\begin{align*}
&\ \Var\left[\int_{\xminh}^{\xmaxh}  \br_p\left(u\right) \Big(\Indicator[x_i\leq x+hu] - F(x+hu) \Big)K(u)f(x+hu)\diff u  \right]\\
&= \iint_{\xminh}^{\xmaxh} \br_p\left(u\right)\br_p\left(v\right)^\Trans K(u)K(v)f(x+hu)f(x+hv)\\
& \qquad\qquad\qquad\qquad\times \left[\int_{\mathbb{R}}\left(\Indicator [t\leq x+hu]-F(x+hu)\right)\left(\Indicator [t\leq x+hv]-F(x+hv)\right) f(t)\diff t\right] \diff u\diff v\\
\tag{I}
&= \iint_{\xminh}^{\xmaxh} \br_p\left(u\right)\br_p\left(v\right)^\Trans K(u)K(v)f(x+hu)f(x+hv) \Big(F(x+h(u\wedge v)) - F(x+hu)F(x+hv)\Big) \diff u\diff v.
\end{align*}
We first consider the interior case, where the above reduces to:
\begin{align*}
&\ \text{(I)}_{\text{interior}} \\
&= \iint_{\mathbb{R}} \br_p\left(u\right)\br_p\left(v\right)^\Trans K(u)K(v)f(x)^2  \Big(F(x) - F(x)^2\Big) \diff u\diff v\\
&\qquad+ h\iint_{\mathbb{R}} (u\wedge v)\br_p\left(u\right)\br_p\left(v\right)^\Trans K(u)K(v) f(x)^3 \diff u\diff v\\
&\qquad- h\iint_{\mathbb{R}} (u+ v)\br_p\left(u\right)\br_p\left(v\right)^\Trans K(u)K(v)f(x)^3 F(x)\diff u\diff v\\
&\qquad + h\iint_{\mathbb{R}} (u+ v)\br_p\left(u\right)\br_p\left(v\right)^\Trans K(u)K(v)f(x)F^{(2)}(x)\Big(F(x) -  F(x)^2\Big) \diff u\diff v + o(h)\\
&=  f(x)^2  \Big(F(x) - F(x)^2\Big)\bS_{p,x}\be_0\be_0^\Trans\bS_{p,x}  \\
&\qquad- h  f(x)^3 F(x) \bS_{p,x}(\be_1\be_0^\Trans + \be_0\be_1^\Trans)\bS_{p,x}\\
&\qquad+ h   f(x)F^{(2)}(x)\Big(F(x) -  F(x)^2\Big) \bS_{p,x}(\be_1\be_0^\Trans + \be_0\be_1^\Trans)\bS_{p,x}\\
&\qquad+ h  f(x)^3 \bGamma_{p,x} + o(h).
\end{align*}
For $x=\xmin+hc$ with $c\in[0, 1)$ in the lower boundary region, 
\begin{align*}
&\ \text{(I)}_{\text{lower boundary}} \\
&=  h\iint_{\mathbb{R}} (u\wedge v+c)\br_p\left(u\right)\br_p\left(v\right)^\Trans K(u)K(v) f(\xmin)^3 \diff u\diff v  + o(h)=  h  f(\xmin)^3 \left(\bGamma_{p,x}+c\bS_{p,x}\be_0\be_0^\Trans\bS_{p,x}\right) + o(h).
\end{align*}
Finally, we have
\begin{align*}
&\ \text{(I)}_{\text{upper boundary}} \\
&=  h\iint_{\mathbb{R}} (u\wedge v -c)\br_p\left(u\right)\br_p\left(v\right)^\Trans K(u)K(v) f(\xmax)^3 \diff u\diff v- h\iint_{\mathbb{R}} (u+ v - 2c)\br_p\left(u\right)\br_p\left(v\right)^\Trans K(u)K(v)f(\xmax)^3 \diff u\diff v + o(h)\\
&=   h f(\xmax)^2 f(\xmax) \left(\bGamma_{p,x}+ c\bS_{p,x}\be_0\be_0^\Trans\bS_{p,x}-\bS_{p,x}(\be_1\be_0^\Trans + \be_0\be_1^\Trans)\bS_{p,x}\right)+o(h).
\end{align*}
With the above results, it is easy to verify the variance formula, provided that we can show the asymptotic normality. 

We first consider the interior case, and verify the Lindeberg condition on the fourth moment. Let $\balpha\in\mathbb{R}^{p+1}$ be an arbitrary nonzero vector, then
\begin{align*}
&\ \sum_i \Expectation\left( \frac{1}{\sqrt{n}} \balpha^\Trans\bN_x(f(x)\bS_{p,x})^{-1}\int_{\xminh}^{\xmaxh}  \br_p\left(u\right) \Big(\Indicator[x_i\leq x+hu] - F(x+hu) \Big)K(u)f(x+hu)\diff u \right)^4\\
&= \frac{1}{n}\Expectation\left(  \balpha^\Trans\bN_x (f(x)\bS_{p,x})^{-1}\int_{\xminh}^{\xmaxh}  \br_p\left(u\right) \Big(\Indicator[x_i\leq x+hu] - F(x+hu) \Big)K(u)f(x+hu)\diff u \right)^4\\
&= \frac{1}{n}\iiiint_{\mathcal{A}} \prod_{j=1,2,3,4} \Big(\balpha^\Trans\bN_x(f(x)\bS_{p,x})^{-1}\br_p\left(u_j\right) K(u_j)\Big) f(x+hu_j) \\
&\qquad \left[\int_{\mathbb{R}} \prod_{j=1,2,3,4}\Big(\Indicator [t\leq x+hu_j]-F(x+hu_j)\Big) f(t)\diff t\right] \diff u_1 \diff u_2 \diff u_3 \diff u_4\\
&\leq \frac{C}{n} \cdot \iiiint_{\mathcal{A}} \prod_{j=1,2,3,4} \Big(\balpha^\Trans\bN_x(f(x)\bS_{p,x})^{-1}\br_p\left(u_j\right) K(u_j)\Big) f(x)\diff u_1 \diff u_2 \diff u_3 \diff u_4 + O\left(\frac{1}{nh}\right),
\end{align*}
where $\mathcal{A}=[\xminh,\xmaxh]^4\subset \mathbb{R}^4$. The first term in the above display is asymptotically negligible, since it is takes the form $C\cdot(\balpha^\Trans\bN_x\be_0)^4/n$ where the constant $C$ depends on the DGP, and is finite. The order of the next term is $1/(nh)$, which comes from multiplying $n^{-1}$, $h^{-2}$ (from the scaling matrix $\bN_x$), and $h$ (from linearization), hence is also negligible.

Under the assumption that $nh\to \infty$, the Lindeberg condition is verified for interior case. The same logic applies to the boundary case, whose proof is easier than the interior case, since the leading term in the calculation is identically zero for $x$ in either the lower or upper boundary. 
\end{proof}

\subsection{Proof of Lemma \ref{lem4: quadratic variance}}

\begin{proof}
For $\hat{\bR}$, we rewrite it as a second order degenerate U-statistic:
\begin{align*}
\hat{\bR} &= \frac{1}{n^2}\sum_{i,j;i< j} \hat{\bU}_{ij},
\end{align*}
where
\begin{align*}
\hat{\bU}_{ij} &= \br_p\left(\frac{x_i-x}{h}\right)\Big(\Indicator[x_j\leq x_i] - F(x_i)\Big)K_h(x_i-x)+ \br_p\left(\frac{x_j-x}{h}\right)\Big(\Indicator[x_i\leq x_j] - F(x_j)\Big)K_h(x_j-x)\\
&- \Expectation\left[ \br_p\left(\frac{x_i-x}{h}\right)\Big(\Indicator[x_j\leq x_i] - F(x_i)\Big)K_h(x_i-x)\Big| x_j \right]- \Expectation\left[ \br_p\left(\frac{x_j-x}{h}\right)\Big(\Indicator[x_i\leq x_j] - F(x_j)\Big)K_h(x_j-x)\Big| x_i \right].
\end{align*}
To compute the leading term, it suffices to consider
\begin{align*}
&\ 2\Expectation\left[ \br_p\left(\frac{x_i-x}{h}\right)\br_p\left(\frac{x_i-x}{h}\right)^\Trans \left(\Indicator [x_j\leq x_i]-F(x_i)\right)^2K_{h}(x_i-x)^2 \right]\\
&= 2\Expectation\left[ \br_p\left(\frac{x_i-x}{h}\right)\br_p\left(\frac{x_i-x}{h}\right)^\Trans \Big( F(x_i) - F(x_i)^2  \Big)K_{h}(x_i-x)^2 \right]\\
&= \frac{2}{h}\int_{\xminh}^{\xmaxh} \br_p\left(v\right)\br_p\left(v\right)^\Trans \Big( F(x+hv) - F(x+hv)^2  \Big) K(v)^2  f(x+ hv)\diff v\\
&= \frac{2}{h}\int_{\xminh}^{\xmaxh} \br_p\left(v\right)\br_p\left(v\right)^\Trans \Big( F(x) - F(x)^2   \Big) K(v)^2  f(x)\diff v + O(1)\\
&=_{\text{interior}} \frac{2}{h}f(x)\left[  F(x) - F(x)^2 \right] \bT_{p,x} + O(1),\\
&=_{\text{boundary}} O(1),
\end{align*}
which closes the proof. 
\end{proof}

\subsection{Proof of Theorem \ref{thm1: asy normality}}

\begin{proof}
This follows from previous lemmas. 
\end{proof}

\subsection{Proof of Theorem \ref{thm2: variance estimation}}

\begin{proof}
First we note that the second half of the theorem follows from the first half and the asymptotic normality result of Theorem \ref{thm1: asy normality}, hence it suffices to prove the first half, i.e. the consistency of $\hat{\mathcal{V}}_{p,v}(x)$. 

The analysis of this estimator is quite involved, since it takes the form of a third order V-statistic. Moreover, since the empirical d.f. $\tilde{F}$ is involved in the formula, a full expansion leads to a fifth order V-statistic. However, some simple tricks will greatly simplify the problem. 

We first split $\hat{\bGamma}_{p,x}$ into four terms, respectively
\begin{align*}
\hat{\bSigma}_{p,x,1} &= \frac{1}{n^3}\sum_{i,j,k} \br_p\left( \frac{x_j-x}{h} \right)\br_p\left( \frac{x_k-x}{h} \right)^\Trans K_h(x_j-x)K_h(x_k-x) \Big(\Indicator[x_i\leq x_j] - F(x_j)\Big) \Big(\Indicator[x_i\leq x_k] - F(x_k)\Big)\\
\hat{\bSigma}_{p,x,2} &= \frac{1}{n^3}\sum_{i,j,k} \br_p\left( \frac{x_j-x}{h} \right)\br_p\left( \frac{x_k-x}{h} \right)^\Trans K_h(x_j-x)K_h(x_k-x)\Big( F(x_j)-\tilde{F}(x_j) \Big)\Big( \Indicator[x_i\leq x_k]-\tilde{F}(x_k) \Big)\\
\hat{\bSigma}_{p,x,3} &= \frac{1}{n^3}\sum_{i,j,k} \br_p\left( \frac{x_j-x}{h} \right)\br_p\left( \frac{x_k-x}{h} \right)^\Trans K_h(x_j-x)K_h(x_k-x)\Big( \Indicator[x_i\leq x_j]-\tilde{F}(x_j) \Big)\Big( F(x_k)-\tilde{F}(x_k) \Big)\\
\hat{\bSigma}_{p,x,4} &= \frac{1}{n^3}\sum_{i,j,k} \br_p\left( \frac{x_j-x}{h} \right)\br_p\left( \frac{x_k-x}{h} \right)^\Trans K_h(x_j-x)K_h(x_k-x) \Big(F(x_j)-\tilde{F}(x_j)\Big) \Big(F(x_k) - \tilde{F}(x_k)\Big).
\end{align*}
Leaving $\hat{\bSigma}_{p,x,1}$ for a while, since it is the key component in this variance estimator. We first consider $\bN_x\hat{\bS}_{p,x}^{-1}\hat{\bSigma}_{p,x,4}\hat{\bS}_{p,x}^{-1}\bN_x$. By the uniform consistency of the empirical d.f., it can be shown easily that
\begin{align*}
\bN_x\hat{\bS}_{p,x}^{-1}\hat{\bSigma}_{p,x,4}\hat{\bS}_{p,x}^{-1}\bN_x &= \Op\left( (nh)^{-1} \right).
\end{align*}
Note that the extra $h^{-1}$ comes from the scaling matrix $\bN_x$, but not the kernel function $K_h$. Next we consider $\bN_x\hat{\bS}_{p,x}^{-1}\hat{\bSigma}_{p,x,2}\hat{\bS}_{p,x}^{-1}\bN_x$, which takes the following form (up to the negligible smoothing bias):
\begin{align*}
\bN_x\hat{\bS}_{p,x}^{-1}\hat{\bSigma}_{p,x,2}\hat{\bS}_{p,x}^{-1}\bN_x =& \bN_x\bH(\bbeta_p(x) - \hat{\bbeta}_p(x)) \left( \frac{1}{n^2}\sum_{i,k} \br_p\left( \frac{x_k-x}{h} \right)^\Trans K_h(x_k-x)\Big( \Indicator[x_i\leq x_k]-\tilde{F}(x_k) \Big) \right)\hat{\bS}_{p,x}^{-1}\bN_x\\
=& \Op((nh)^{-1/2}) = \op(1),
\end{align*}
where the last line uses the asymptotic normality of $\hat{\bbeta}_p(x)$. For $\hat{\bSigma}_{p,x,1}$, we make the observation that it is possible to ignore all ``diagonal'' terms, meaning that 
\begin{align*}
\hat{\bSigma}_{p,x,1} &= \frac{1}{n^3}\sum_{\substack{i,j,k \\ \text{distinct}}} \br_p\left( \frac{x_j-x}{h} \right)\br_p\left( \frac{x_k-x}{h} \right)^\Trans K_h(x_j-x)K_h(x_k-x) \Big(\Indicator[x_i\leq x_j] - F(x_j)\Big) \Big(\Indicator[x_i\leq x_k] - F(x_k)\Big) + \op(h),
\end{align*}
under the assumption that $nh^2\to \infty$. As a surrogate, define 
\begin{align*}
\bU_{i,j,k} &= \br_p\left( \frac{x_j-x}{h} \right)\br_p\left( \frac{x_k-x}{h} \right)^\Trans K_h(x_j-x)K_h(x_k-x)\Big(\Indicator[x_i\leq x_j] - F(x_j)\Big) \Big(\Indicator[x_i\leq x_k] - F(x_k)\Big),
\end{align*}
which means
\begin{align*}
\hat{\bSigma}_{p,x,1} &= \frac{1}{n^3}\sum_{\substack{i,j,k \\ \text{distinct}}} \bU_{i,j,k}. 
\end{align*}
The critical step is to further decompose the above into
\begin{align*}
\tag{I}\hat{\bSigma}_{p,x,1} &= \frac{1}{n^3}\sum_{\substack{i,j,k \\ \text{distinct}}} \Expectation[\bU_{i,j,k}|x_i] \\
\tag{II}&+ \frac{1}{n^3}\sum_{\substack{i,j,k \\ \text{distinct}}} \Big(\bU_{i,j,k} - \Expectation[\bU_{i,j,k}|x_i,x_j]\Big)\\
\tag{III}&+ \frac{1}{n^3}\sum_{\substack{i,j,k \\ \text{distinct}}} \Big(\Expectation[\bU_{i,j,k}|x_i,x_j] - \Expectation[\bU_{i,j,k}|x_i]\Big).
\end{align*}
We already investigated the properties of term (I) in Lemma \ref{lem3: linear variance}, hence it remains to show that both (II) and (III) are $o(h)$, hence does not affect the estimation of asymptotic variance. We consider (II) as an example, and the analysis of (III) is similar. Since (II) has zero expectation, we consider its variance (for simplicity treat $\bU$ as a scaler):
\begin{align*}
\Var[\text{(II)}] &= \Expectation\left[ \frac{1}{n^6}\sum_{\substack{i,j,k \\ \text{distinct}}}\sum_{\substack{i',j',k' \\ \text{distinct}}} \Big(\bU_{i,j,k} - \Expectation[\bU_{i,j,k}|x_i,x_j]\Big)\Big(\bU_{i,j,k} - \Expectation[\bU_{i',j',k'}|x_{i'},x_{j'}]\Big) \right].
\end{align*}
The expectation will be zero if the six indices are all distinct. Similarly, when there are only two indices among the six are equal, the expectation will be zero \textit{unless} $k=k'$, hence
\begin{align*}
\Var[\text{(II)}] &= \Expectation\left[ \frac{1}{n^6}\sum_{\substack{i,j,k \\ \text{distinct}}}\sum_{\substack{i',j',k' \\ \text{distinct}}} \Big(\bU_{i,j,k} - \Expectation[\bU_{i,j,k}|x_i,x_j]\Big)\Big(\bU_{i,j,k} - \Expectation[\bU_{i',j',k'}|x_{i'},x_{j'}]\Big) \right]\\
&= \Expectation\left[ \frac{1}{n^6}\sum_{\substack{i,j,k,i'j' \\ \text{distinct}}} \Big(\bU_{i,j,k} - \Expectation[\bU_{i,j,k}|x_i,x_j]\Big)\Big(\bU_{i,j,k} - \Expectation[\bU_{i',j',k}|x_{i'},x_{j'}]\Big) \right]\\
&+ \cdots,
\end{align*}
where $\cdots$ represent cases where more than two indices among the six are equal. We can easily compute the order from the above as
\begin{align*}
\Var[\text{(II)}] &= O(n^{-1}) + O((nh)^{-2}),
\end{align*}
which shows that
\begin{align*}
\text{(II)} &= \Op(n^{-1/2} + (nh)^{-1}) = \op(h),
\end{align*}
which closes the proof. 
\end{proof}

\subsection{Proof of Lemma \ref{lem5: higher order bias}}

\begin{proof}
We rely on Lemma \ref{lem1: denominator} and \ref{lem2: smoothing bias} (note that whether the weights are estimated is irrelevant here), hence will not repeat arguments already established there. Instead, extra care will be given to ensure the characterization of higher order bias. 

Consider the case where with enough smoothness on $G$, then the bias is characterized by
\begin{align*}
&\ h^{-v}v!\be_v^\Trans\left[f(x)\bS_{p,x} + hF^{(2)}(x)\tilde{\bS}_{p,x} + o(h) + \Op(1/\sqrt{nh}) \right]^{-1} \\ 
&\ \qquad \left[ h^{p+1}\frac{F^{(p+1)}(x)}{(p+1)!}f(x)\bc_{p,x}+h^{p+2}\left[\frac{F^{(p+2)}(x)}{(p+2)!}f(x)+\frac{F^{(p+1)}(x)}{(p+1)!}F^{(2)}(x)\right]\tilde{\bc}_{p,x}  + o(h^{p+2})\right]\\
&= h^{-v}v!\be_v^\Trans\left[ \frac{1}{f(x)}\bS_{p,x}^{-1}-h\frac{F^{(2)}(x)}{[f(x)]^2}\bS_{p,x}^{-1}\tilde \bS_{p,x} \bS_{p,x}^{-1}+\Op\left(1/\sqrt{nh}\right)  \right]\\
&\ \qquad \left[ h^{p+1}\frac{F^{(p+1)}(x)}{(p+1)!}f(x)\bc_{p,x}+h^{p+2}\left[\frac{F^{(p+2)}(x)}{(p+2)!}f(x)+\frac{F^{(p+1)}(x)}{(p+1)!}F^{(2)}(x)\right]\tilde{\bc}_{p,x}  + o(h^{p+2})\right]\{1+\op(1)\},
\end{align*}
which gives the desired result. Here $\tilde{\bS}_{p,x} = \int_{\frac{\xmin-x}{h}}^{\frac{\xmax-x}{h}} u\br_p(u)\br_p(u)^\Trans k(u)\diff u$.
And for the last line to hold, one needs the extra condition $nh^3\to \infty$ so that $\Op\left(1/\sqrt{nh}\right)=\op(h)$. See \cite{Fan-Gijbels_1996_Book} (Theorem 3.1, pp. 62). 
\end{proof}

\subsection{Proof of Lemma \ref{lem6: bias estimation}}

\begin{proof}
The proof resembles that of Lemma \ref{lem1: denominator}, and is omitted here. 
\end{proof}

\subsection{Proof of Theorem \ref{thm3: bandwidth consistency}}

\begin{proof}
The proof splits into two cases. We sketch one of them. Assume either $x$ is boundary or $p-v$ is odd, the  MSE-optimal bandwidth is asymptotically equivalent to the following:
\begin{align*}
\frac{\tilde{h}_{ p,v}(x)}{{h}_{ p,v}(x)}\to 1,\qquad \tilde{h}_{ p,v}(x) = \left(\frac{1}{n} \frac{(2v-1)f(x)\be_v^\Trans\bS_{p,x}^{-1}\bGamma_{p,x}\bS_{p,x}^{-1}\be_v}{(2p-2v+2)(\frac{F^{(p+1)}(x)}{(p+1)!}\be_v^\Trans \bS_{p,x}^{-1}\bc_{p,x} )^2} \right)^{\frac{1}{2p+1}},
\end{align*}
which is obtained by optimizing  MSE ignoring the higher order bias term. With consistency of the preliminary estimates, it can be shown that 
\begin{align*}
\hat{h}_{ p,v}(x) &= \left(\frac{1}{n} \frac{(2v-1)\hat{\sigma}_{p,v}(x)^2n\ell^{2v-1}}{(2p-2v+2)(v!\frac{\hat{F}^{(p+1)}(x)}{(p+1)!}\be_v^\Trans \bS_{p,x}^{-1}\bc_{p,x} )^2} \right)^{\frac{1}{2p+1}}\{1+\op(1)\}.
\end{align*}
Apply the consistency assumption of the preliminary estimates again, one can easily show that $\hat{h}_{ p,v}(x)$ is consistent both in rate and constant. 

A similar argument can be made for the other case, and is omitted here. 
\end{proof}

\subsection{Proof of Lemma \ref{lem7: denominator, unrestricted}} 

\begin{proof}
This resembles the proof of Lemma \ref{lem1: denominator}, and we only perform the mean computation. To start, 
\begin{align*}
&\ \Expectation\left[ \frac{1}{n}\bX_h' \bK_h \bX_h \right] = \Expectation\left[ \br_p\left(\frac{x_i-\cutoff}{h}\right)\br_p\left(\frac{x_i-\cutoff}{h}\right)'\frac{1}{h}K\left(\frac{x_i-\cutoff}{h}\right) \right]\\
&= \Expectation\left[\left. \br_p\left(\frac{x_i-\cutoff}{h}\right)\br_p\left(\frac{x_i-\cutoff}{h}\right)'\frac{1}{h}K\left(\frac{x_i-\cutoff}{h}\right) \right| x_i<\cutoff\right]F(\cutoff)\\
&\qquad + \Expectation\left[\left. \br_p\left(\frac{x_i-\cutoff}{h}\right)\br_p\left(\frac{x_i-\cutoff}{h}\right)'\frac{1}{h}K\left(\frac{x_i-\cutoff}{h}\right) \right| x_i\geq\cutoff\right](1-F(\cutoff)).
\end{align*}
Then by Lemma \ref{lem1: denominator}, the first term takes the form:
\begin{align*}
&\ \Expectation\left[\left. \br_p\left(\frac{x_i-\cutoff}{h}\right)\br_p\left(\frac{x_i-\cutoff}{h}\right)'\frac{1}{h}K\left(\frac{x_i-\cutoff}{h}\right) \right| x_i<\cutoff\right]F(\cutoff)\\
&= f(\cutoff- |x_i<\cutoff)F(\cutoff) \int_{-1}^0 \br_{-,p}(u)\br_{-,p}(u)'K(u)\diff u + O(h),
\end{align*}
where $f(\cutoff- |x_i<\cutoff)$ is the one-sided density of $x_i$ at the cutoff, conditional on $x_i<\cutoff$. Alternatively, we can simplify by the fact that $f(\cutoff |x_i<\cutoff)F(\cutoff) = f(\cutoff -)$. Similarly, one has
\begin{align*}
&\ \Expectation\left[\left. \br_p\left(\frac{x_i-\cutoff}{h}\right)\br_p\left(\frac{x_i-\cutoff}{h}\right)'\frac{1}{h}K\left(\frac{x_i-\cutoff}{h}\right) \right| x_i\geq \cutoff\right](1-F(\cutoff))\\
&= f(\cutoff+ |x_i\geq\cutoff)(1-F(\cutoff)) \int_{0}^1 \br_{+,p}(u)\br_{+,p}(u)'K(u)\diff u + O(h),
\end{align*}
and that $f(\cutoff+ |x_i\geq\cutoff)(1-F(\cutoff)) = f(\cutoff+ )$. The rest of the proof follows standard variance calculation, and is not repeated here. 
\end{proof}

\subsection{Proof of Lemma \ref{lem8: smoothing bias, unrestricted}}

\begin{proof}
This follows from Lemma \ref{lem2: smoothing bias} by splitting the bias calculation for the two subsamples, below and above the cutoff $\cutoff$. 
\end{proof}

\subsection{Proof of Lemma \ref{lem9: linear variance, unrestricted}}

\begin{proof}
To start,
\begin{align*}
&\ \int_{-1}^{1}  \br_p\left(u\right)\Big(\tilde{F}(\cutoff+hu) - F(\cutoff+hu) \Big)K(u)f(\cutoff+hu)\diff u = \frac{1}{n}\int_{-1}^{1}  \br_p\left(u\right) \Big(\Indicator[x_i\leq \cutoff+hu] - F(\cutoff+hu) \Big)K(u)f(\cutoff+hu)\diff u,
\end{align*}
and
\begin{align*}
&\ \Var\left[\int_{-1}^{1}  \br_p\left(u\right) \Big(\Indicator[x_i\leq \cutoff+hu] - F(\cutoff+hu) \Big)K(u)f(\cutoff+hu)\diff u  \right]\\
&= \iint_{-1}^{1} \br_p\left(u\right)\br_p\left(v\right)^\Trans K(u)K(v)f(\cutoff+hu)f(\cutoff+hv)\\ 
&\qquad\qquad\qquad\qquad\times\left[\int_{\mathbb{R}}\left(\Indicator [t\leq \cutoff+hu]-F(\cutoff+hu)\right)\left(\Indicator [t\leq \cutoff+hv]-F(\cutoff+hv)\right) f(t)\diff t\right] \diff u\diff v\\
\tag{I}
&= \iint_{-1}^{1} \br_p\left(u\right)\br_p\left(v\right)^\Trans K(u)K(v)f(\cutoff+hu)f(\cutoff+hv) \Big(F(\cutoff+h(u\wedge v)) - F(\cutoff+hu)F(\cutoff+hv)\Big) \diff u\diff v.
\end{align*}
Now we split the integral of (I) into four regions. 
\begin{align*}
(u<0,v<0)\ \text{(I)} &= \iint_{-1}^{0} \br_{-,p}\left(u\right)\br_{-,p}\left(v\right)^\Trans K(u)K(v)f(\cutoff+hu)f(\cutoff+hv) \Big(F(\cutoff+h(u\wedge v)) - F(\cutoff+hu)F(\cutoff+hv)\Big) \diff u\diff v\\
&=  f(\cutoff-)^2  \Big(F(\cutoff) - F(\cutoff)^2\Big)\bS_{-,p}\be_{0,-}\be_{0,-}^\Trans\bS_{-,p}  \\
&\qquad- h  f(\cutoff-)^3 F(\cutoff) \bS_{-,p}(\be_{1,-}\be_{0,-}^\Trans + \be_{0,-}\be_{1,-}^\Trans)\bS_{-,p}\\
&\qquad+ h   f(\cutoff-)F^{(2)}(\cutoff-)\Big(F(\cutoff) -  F(\cutoff)^2\Big) \bS_{-,p}(\be_{1,-}\be_{0,-}^\Trans + \be_{0,-}\be_{1,-}^\Trans)\bS_{-,p}\\
&\qquad+ h  f(\cutoff-)^3 \bGamma_{-,p} + O(h^2),
\end{align*}
and
\begin{align*}
(u\geq 0,v\geq 0)\ \text{(I)} &= \iint_{0}^{1} \br_{+,p}\left(u\right)\br_{+,p}\left(v\right)^\Trans K(u)K(v)f(\cutoff+hu)f(\cutoff+hv) \Big(F(\cutoff+h(u\wedge v)) - F(\cutoff+hu)F(\cutoff+hv)\Big) \diff u\diff v\\
&=  f(\cutoff+)^2  \Big(F(\cutoff) - F(\cutoff)^2\Big)\bS_{+,p}\be_{0,+}\be_{0,+}^\Trans\bS_{+,p}  \\
&\qquad- h  f(\cutoff+)^3 F(\cutoff) \bS_{+,p}(\be_{1,+}\be_{0,+}^\Trans + \be_{0,+}\be_{1,+}^\Trans)\bS_{+,p}\\
&\qquad+ h   f(\cutoff+)F^{(2)}(\cutoff+)\Big(F(\cutoff) -  F(\cutoff)^2\Big) \bS_{+,p}(\be_{1,+}\be_{0,+}^\Trans + \be_{0,+}\be_{1,+}^\Trans)\bS_{+,p}\\
&\qquad+ h  f(\cutoff+)^3 \bGamma_{+,p} + O(h^2),
\end{align*}
and
\begin{align*}
(u<0,v\geq 0)\ \text{(I)} &= \iint_{[-1,0]\times[0,1]} \br_{-,p}\left(u\right)\br_{+,p}\left(v\right)^\Trans K(u)K(v)f(\cutoff+hu)f(\cutoff+hv) F(\cutoff+hu) \Big(1- F(\cutoff+hv)\Big) \diff u\diff v\\
&= \left[\int_{-1}^0 \br_{-,p}\left(u\right) K(u)f(\cutoff+hu)F(\cutoff+hu)  \diff u\right]\left[\int_{0}^1 \br_{+,p}\left(v\right)^\Trans K(v)f(\cutoff+hv)  \Big(1- F(\cutoff+hv)\Big) \diff v\right]\\
&= \Big[ f(\cutoff-)F(\cutoff) \bS_{-,p}\be_{0,-} + h\Big( f(\cutoff-)^2 + F^{(2)}(\cutoff-)F(\cutoff) \Big) \bS_{-,p}\be_{1,-} + O(h^2)\Big]\\
&\qquad \qquad \Big[ f(\cutoff+)(1-F(\cutoff)) \bS_{+,p}\be_{0,+} + h\Big( -f(\cutoff+)^2 + F^{(2)}(\cutoff+)(1-F(\cutoff)) \Big) \bS_{+,p}\be_{1,+} + O(h^2)\Big]',
\end{align*}
and
\begin{align*}
(u\geq 0,v< 0)\ \text{(I)} &= \iint_{[0,1]\times[-1,0]} \br_{-,p}\left(u\right)\br_{+,p}\left(v\right)^\Trans K(u)K(v)f(\cutoff+hu)f(\cutoff+hv) F(\cutoff+hv) \Big(1- F(\cutoff+hu)\Big) \diff u\diff v\\
&= \left[\int_{0}^1 \br_{+,p}\left(u\right) K(u)f(\cutoff+hu)(1-F(\cutoff+hu))  \diff u\right]\left[\int_{-1}^{0} \br_{-,p}\left(v\right)^\Trans K(v)f(\cutoff+hv)  F(\cutoff+hv)\Big) \diff v\right]\\
&= \Big[ f(\cutoff+)(1-F(\cutoff)) \bS_{+,p}\be_{0,+} + h\Big( -f(\cutoff+)^2 + F^{(2)}(\cutoff+)(1-F(\cutoff)) \Big) \bS_{+,p}\be_{1,+} + O(h^2)\Big]\\
&\qquad \qquad \Big[ f(\cutoff-)F(\cutoff-) \bS_{-,p}\be_{0,-} + h\Big( f(\cutoff-)^2 + F^{(2)}(\cutoff-)F(\cutoff) \Big) \bS_{-,p}\be_{1,-} + O(h^2)\Big]'.
\end{align*}
Let $\bS_{-,p}^{-1}$ and $\bS_{+,p}^{-1}$ be the Moore–Penrose inverse of $\bS_{-,p}$ and $\bS_{+,p}$, respectively. Then
\begin{align*}
&\ \Var\left[ (\be_{1,+}-\be_{1,-})'\sqrt{\frac{n}{h}}(f(\cutoff+)\bS_{+,p}+f(\cutoff-)\bS_{-,p})^{-1}\hat{\bL} \right]\\
&= f(\cutoff-) \be_{1,-}'\bS_{-,p}^{-1}\bGamma_{-,p}\bS_{-,p}^{-1}\be_{1,-} + f(\cutoff+) \be_{1,+}'\bS_{+,p}^{-1}\bGamma_{+,p}\bS_{+,p}^{-1}\be_{1,+} + O(h).
\end{align*}
\end{proof}

\subsection{Proof of Lemma \ref{lem10: quadratic variance, unrestricted}}

\begin{proof}
This follows from Lemma \ref{lem4: quadratic variance} by splitting the bias calculation for the two subsamples, below and above the cutoff $\cutoff$. 
\end{proof}

\subsection{Proof of Corollary \ref{coro1: asy normal, unrestricted}}

\begin{proof}
This follows from the previous lemmas and verifying the Lindeberg condition. See also the proof of Lemma \ref{lem3: linear variance}, Theorem \ref{thm1: asy normality} and Theorem \ref{thm2: variance estimation}.
\end{proof}

\subsection{Proof of Lemma \ref{lem11: denominator restricted}}

\begin{proof}
This follows from Lemma \ref{lem1: denominator} by splitting the bias calculation for the two subsamples, below and above the cutoff $\cutoff$. See also the proof of Lemma \ref{lem7: denominator, unrestricted}. 
\end{proof}

\subsection{Proof of Lemma \ref{lem12: smoothing bias, restricted}}

\begin{proof}
This follows from Lemma \ref{lem2: smoothing bias} by splitting the bias calculation for the two subsamples, below and above the cutoff $\cutoff$. 
\end{proof}

\subsection{Proof of Lemma \ref{lem13: linear variance, restricted}}

\begin{proof}
To start,
\begin{align*}
&\ \int_{-1}^{1}  \br_p\left(u\right)\Big(\tilde{F}(\cutoff+hu) - F(\cutoff+hu) \Big)K(u)f(\cutoff+hu)\diff u = \frac{1}{n}\int_{-1}^{1}  \br_p\left(u\right) \Big(\Indicator[x_i\leq \cutoff+hu] - F(\cutoff+hu) \Big)K(u)f(\cutoff+hu)\diff u,
\end{align*}
and
\begin{align*}
&\ \Var\left[\int_{-1}^{1}  \br_p\left(u\right) \Big(\Indicator[x_i\leq \cutoff+hu] - F(\cutoff+hu) \Big)K(u)f(\cutoff+hu)\diff u  \right]\\
&= \iint_{-1}^{1} \br_p\left(u\right)\br_p\left(v\right)^\Trans K(u)K(v)f(\cutoff+hu)f(\cutoff+hv)\\
&\qquad\qquad\qquad\qquad\times\left[\int_{\mathbb{R}}\left(\Indicator [t\leq \cutoff+hu]-F(\cutoff+hu)\right)\left(\Indicator [t\leq \cutoff+hv]-F(\cutoff+hv)\right) f(t)\diff t\right] \diff u\diff v\\
\tag{I}&= \iint_{-1}^{1} \br_p\left(u\right)\br_p\left(v\right)^\Trans K(u)K(v)f(\cutoff+hu)f(\cutoff+hv) \Big(F(\cutoff+h(u\wedge v)) - F(\cutoff+hu)F(\cutoff+hv)\Big) \diff u\diff v.
\end{align*}
Now we split the integral of (I) into four regions. 
\begin{align*}
(u<0,v<0)\ \text{(I)} &= \iint_{-1}^{0} \br_{-,p}\left(u\right)\br_{-,p}\left(v\right)^\Trans K(u)K(v)f(\cutoff+hu)f(\cutoff+hv) \Big(F(\cutoff+h(u\wedge v)) - F(\cutoff+hu)F(\cutoff+hv)\Big) \diff u\diff v\\
&=  f(\cutoff-)^2  \Big(F(\cutoff) - F(\cutoff)^2\Big)\bS_{-,p}\be_{0}\be_{0}^\Trans\bS_{-,p}  \\
&\qquad- h  f(\cutoff-)^3 F(\cutoff) \bS_{-,p}(\be_{1,-}\be_{0}^\Trans + \be_{0}\be_{1,-}^\Trans)\bS_{-,p}\\
&\qquad+ h   f(\cutoff-)F^{(2)}(\cutoff)\Big(F(\cutoff) -  F(\cutoff)^2\Big) \bS_{-,p}(\be_{1,-}\be_{0}^\Trans + \be_{0}\be_{1,-}^\Trans)\bS_{-,p}\\
&\qquad+ h  f(\cutoff-)^3 \bGamma_{-,p} + O(h^2),
\end{align*}
and
\begin{align*}
(u\geq 0,v\geq 0)\ \text{(I)} &= \iint_{0}^{1} \br_{+,p}\left(u\right)\br_{+,p}\left(v\right)^\Trans K(u)K(v)f(\cutoff+hu)f(\cutoff+hv) \Big(F(\cutoff+h(u\wedge v)) - F(\cutoff+hu)F(\cutoff+hv)\Big) \diff u\diff v\\
&=  f(\cutoff+)^2  \Big(F(\cutoff) - F(\cutoff)^2\Big)\bS_{+,p}\be_{0}\be_{0}^\Trans\bS_{+,p}  \\
&\qquad- h  f(\cutoff+)^3 F(\cutoff) \bS_{+,p}(\be_{1,+}\be_{0}^\Trans + \be_{0}\be_{1,+}^\Trans)\bS_{+,p}\\
&\qquad+ h   f(\cutoff+)F^{(2)}(\cutoff)\Big(F(\cutoff) -  F(\cutoff)^2\Big) \bS_{+,p}(\be_{1,+}\be_{0}^\Trans + \be_{0}\be_{1,+}^\Trans)\bS_{+,p}\\
&\qquad+ h  f(\cutoff+)^3 \bGamma_{+,p} + O(h^2),
\end{align*}
and
\begin{align*}
(u<0,v\geq 0)\ \text{(I)} &= \iint_{[-1,0]\times[0,1]} \br_{-,p}\left(u\right)\br_{+,p}\left(v\right)^\Trans K(u)K(v)f(\cutoff+hu)f(\cutoff+hv) F(\cutoff+hu) \Big(1- F(\cutoff+hv)\Big) \diff u\diff v\\
&= \left[\int_{-1}^0 \br_{-,p}\left(u\right) K(u)f(\cutoff+hu)F(\cutoff+hu)  \diff u\right]\left[\int_{0}^1 \br_{+,p}\left(v\right)^\Trans K(v)f(\cutoff+hv)  \Big(1- F(\cutoff+hv)\Big) \diff v\right]\\
&= \Big[ f(\cutoff-)F(\cutoff) \bS_{-,p}\be_{0} + h\Big( f(\cutoff-)^2 + F^{(2)}(\cutoff)F(\cutoff) \Big) \bS_{-,p}\be_{1,-} + O(h^2)\Big]\\
&\qquad \qquad \Big[ f(\cutoff+)(1-F(\cutoff)) \bS_{+,p}\be_{0} + h\Big( -f(\cutoff+)^2 + F^{(2)}(\cutoff)(1-F(\cutoff)) \Big) \bS_{+,p}\be_{1,+} + O(h^2)\Big]',
\end{align*}
and
\begin{align*}
(u\geq 0,v< 0)\ \text{(I)} &= \iint_{[0,1]\times[-1,0]} \br_{-,p}\left(u\right)\br_{+,p}\left(v\right)^\Trans K(u)K(v)f(\cutoff+hu)f(\cutoff+hv) F(\cutoff+hv) \Big(1- F(\cutoff+hu)\Big) \diff u\diff v\\
&= \left[\int_{0}^1 \br_{+,p}\left(u\right) K(u)f(\cutoff+hu)(1-F(\cutoff+hu))  \diff u\right]\left[\int_{-1}^{0} \br_{-,p}\left(v\right)^\Trans K(v)f(\cutoff+hv)  F(\cutoff+hv)\Big) \diff v\right]\\
&= \Big[ f(\cutoff+)(1-F(\cutoff)) \bS_{+,p}\be_{0} + h\Big( -f(\cutoff+)^2 + F^{(2)}(\cutoff)(1-F(\cutoff)) \Big) \bS_{+,p}\be_{1,+} + O(h^2)\Big]\\
&\qquad \qquad \Big[ f(\cutoff-)F(\cutoff) \bS_{-,p}\be_{0} + h\Big( f(\cutoff-)^2 + F^{(2)}(\cutoff)F(\cutoff) \Big) \bS_{-,p}\be_{1,-} + O(h^2)\Big]'.
\end{align*}

By collecting terms, one has
\begin{align*}
\text{(I)} &= \Big(f(\cutoff+)\bS_{+,p} + f(\cutoff+)\bS_{-,p}\Big)\be_{0}\be_{0}'\Big(f(\cutoff+)\bS_{+,p} + f(\cutoff+)\bS_{-,p}\Big)'\\
&  \qquad- h  f(\cutoff-) F(\cutoff) f(\cutoff-)\bS_{-,p}\be_{1,-}\be_{0}^\Trans (f(\cutoff+)\bS_{+,p}+f(\cutoff-)\bS_{-,p})\\
&  \qquad+ h   \frac{F^{(2)}(\cutoff)}{f(\cutoff-)}F(\cutoff)(1 -  F(\cutoff)) f(\cutoff-)\bS_{-,p}\be_{1,-}\be_{0}^\Trans (f(\cutoff+)\bS_{+,p}+f(\cutoff-)\bS_{-,p})\\
&  \qquad- h  f(\cutoff-) F(\cutoff) (f(\cutoff+)\bS_{+,p}+f(\cutoff-)\bS_{-,p})\be_{0}\be_{1,-}^\Trans f(\cutoff-)\bS_{-,p}\\
&  \qquad+h  \frac{F^{(2)}(\cutoff)}{f(\cutoff-)}F(\cutoff)   (1-F(\cutoff)) (f(\cutoff+)\bS_{+,p}+f(\cutoff-)\bS_{-,p})\be_{0}\be_{1,-}^\Trans f(\cutoff-)\bS_{-,p}\\
&  \qquad- h  f(\cutoff+) F(\cutoff) f(\cutoff+)\bS_{+,p}\be_{1,+}\be_{0}^\Trans(f(\cutoff+)\bS_{+,p}+f(\cutoff-)\bS_{-,p})\\
&  \qquad+h \frac{F^{(2)}(\cutoff)}{f(\cutoff+)}(1-F(\cutoff))F(\cutoff) f(\cutoff+)\bS_{+,p}\be_{1,+}\be_{0}^\Trans (f(\cutoff+)\bS_{+,p}+f(\cutoff-)\bS_{-,p})\\
&  \qquad- h  f(\cutoff+) F(\cutoff) (f(\cutoff+)\bS_{+,p}+f(\cutoff-)\bS_{-,p})\be_{0}\be_{1,+}^\Trans f(\cutoff+)\bS_{+,p}\\
&  \qquad+h \frac{F^{(2)}(\cutoff)}{f(\cutoff+)}F(\cutoff)(1-F(\cutoff))  (f(\cutoff+)\bS_{+,p}+f(\cutoff-)\bS_{-,p})\be_{0}\be_{1,+}^\Trans f(\cutoff+)\bS_{+,p}\\
&  \qquad+hf(\cutoff-)  f(\cutoff-)\bS_{-,p}\be_{1,-}\be_{0}^\Trans f(\cutoff+)\bS_{+,p}\\
&  \qquad+h  f(\cutoff-)  f(\cutoff+)\bS_{+,p}\be_{0}\be_{1,-}^\Trans f(\cutoff-)\bS_{-,p}\\
&  \qquad+h  (f(\cutoff+)^3 \bGamma_{+,p}+  f(\cutoff-)^3 \bGamma_{-,p}).
\end{align*}
Next, we note that 
\begin{align*}
\bS_{+,p}\be_{1,-} = \bS_{-,p}\be_{1,+} = \mathbf{0},
\end{align*}
which implies
\begin{align*}
\text{(I)} &= \Big(f(\cutoff+)\bS_{+,p} + f(\cutoff+)\bS_{-,p}\Big)\be_{0}\be_{0}'\Big(f(\cutoff+)\bS_{+,p} + f(\cutoff+)\bS_{-,p}\Big)'\\
&  \qquad- h  f(\cutoff-) F(\cutoff) (f(\cutoff+)\bS_{+,p}+f(\cutoff-)\bS_{-,p})\be_{1,-}\be_{0}^\Trans (f(\cutoff+)\bS_{+,p}+f(\cutoff-)\bS_{-,p})\\
&  \qquad+ h   \frac{F^{(2)}(\cutoff)}{f(\cutoff-)}F(\cutoff)(1 -  F(\cutoff)) (f(\cutoff+)\bS_{+,p}+f(\cutoff-)\bS_{-,p})\be_{1,-}\be_{0}^\Trans (f(\cutoff+)\bS_{+,p}+f(\cutoff-)\bS_{-,p})\\
&  \qquad- h  f(\cutoff-) F(\cutoff) (f(\cutoff+)\bS_{+,p}+f(\cutoff-)\bS_{-,p})\be_{0}\be_{1,-}^\Trans (f(\cutoff+)\bS_{+,p}+f(\cutoff-)\bS_{-,p})\\
&  \qquad+h  \frac{F^{(2)}(\cutoff)}{f(\cutoff-)}F(\cutoff)   (1-F(\cutoff)) (f(\cutoff+)\bS_{+,p}+f(\cutoff-)\bS_{-,p})\be_{0}\be_{1,-}^\Trans (f(\cutoff+)\bS_{+,p}+f(\cutoff-)\bS_{-,p})\\
&  \qquad- h  f(\cutoff+) F(\cutoff) (f(\cutoff+)\bS_{+,p}+f(\cutoff-)\bS_{-,p})\be_{1,+}\be_{0}^\Trans(f(\cutoff+)\bS_{+,p}+f(\cutoff-)\bS_{-,p})\\
&  \qquad+h \frac{F^{(2)}(\cutoff)}{f(\cutoff+)}(1-F(\cutoff))F(\cutoff) (f(\cutoff+)\bS_{+,p}+f(\cutoff-)\bS_{-,p})\be_{1,+}\be_{0}^\Trans (f(\cutoff+)\bS_{+,p}+f(\cutoff-)\bS_{-,p})\\
&  \qquad- h  f(\cutoff+) F(\cutoff) (f(\cutoff+)\bS_{+,p}+f(\cutoff-)\bS_{-,p})\be_{0}\be_{1,+}^\Trans (f(\cutoff+)\bS_{+,p}+f(\cutoff-)\bS_{-,p})\\
&  \qquad+h \frac{F^{(2)}(\cutoff)}{f(\cutoff+)}F(\cutoff)(1-F(\cutoff))  (f(\cutoff+)\bS_{+,p}+f(\cutoff-)\bS_{-,p})\be_{0}\be_{1,+}^\Trans (f(\cutoff+)\bS_{+,p}+f(\cutoff-)\bS_{-,p})\\
&  \qquad+hf(\cutoff-)  (f(\cutoff+)\bS_{+,p}+f(\cutoff-)\bS_{-,p})\be_{1,-}\be_{0}^\Trans f(\cutoff+)\bS_{+,p}\\
&  \qquad+h  f(\cutoff-)  f(\cutoff+)\bS_{+,p}\be_{0}\be_{1,-}^\Trans (f(\cutoff+)\bS_{+,p}+f(\cutoff-)\bS_{-,p})\\
&  \qquad+h  (f(\cutoff+)^3 \bGamma_{+,p}+  f(\cutoff-)^3 \bGamma_{-,p}).
\end{align*}
Next note that
\begin{align*}
\bGamma_{-,p} &= \iint_{[-1,0]^2} (u\wedge v)\br_{-,p}(u)\br_{-,p}(v)'K(u)K(v) dudv\\
&= \iint_{[0,1]^2} ((-u)\wedge (-v))\br_{-,p}(-u)\br_{-,p}(-v)'K(u)K(v) dudv\\
&= \iint_{[0,1]^2} (u\wedge v - u - v)\boldsymbol{\Psi}\br_{+,p}(u)\br_{+,p}(v)'\boldsymbol{\Psi}K(u)K(v) dudv\\
&= \boldsymbol{\Psi}\bGamma_{+,p}\boldsymbol{\Psi} - \boldsymbol{\Psi}\bS_{+,p}\be_{1,+}\be_{0}'\bS_{+,p}\boldsymbol{\Psi}-\boldsymbol{\Psi}\bS_{+,p}\be_{0}\be_{1,+}'\bS_{+,p}\boldsymbol{\Psi}\\
&= \boldsymbol{\Psi}\bGamma_{+,p}\boldsymbol{\Psi} + \bS_{-,p}\be_{1,-}\be_{0}'\bS_{-,p}+\bS_{-,p}\be_{0}\be_{1,-}'\bS_{-,p},
\end{align*}
then
\begin{align*}
\text{(I)} &= \Big(f(\cutoff+)\bS_{+,p} + f(\cutoff+)\bS_{-,p}\Big)\be_{0}\be_{0}'\Big(f(\cutoff+)\bS_{+,p} + f(\cutoff+)\bS_{-,p}\Big)'\\
&  \qquad- h  f(\cutoff-) F(\cutoff) (f(\cutoff+)\bS_{+,p}+f(\cutoff-)\bS_{-,p})\be_{1,-}\be_{0}^\Trans (f(\cutoff+)\bS_{+,p}+f(\cutoff-)\bS_{-,p})\\
&  \qquad+ h   \frac{F^{(2)}(\cutoff)}{f(\cutoff-)}F(\cutoff)(1 -  F(\cutoff)) (f(\cutoff+)\bS_{+,p}+f(\cutoff-)\bS_{-,p})\be_{1,-}\be_{0}^\Trans (f(\cutoff+)\bS_{+,p}+f(\cutoff-)\bS_{-,p})\\
&  \qquad- h  f(\cutoff-) F(\cutoff) (f(\cutoff+)\bS_{+,p}+f(\cutoff-)\bS_{-,p})\be_{0}\be_{1,-}^\Trans (f(\cutoff+)\bS_{+,p}+f(\cutoff-)\bS_{-,p})\\
&  \qquad+h  \frac{F^{(2)}(\cutoff)}{f(\cutoff-)}F(\cutoff)   (1-F(\cutoff)) (f(\cutoff+)\bS_{+,p}+f(\cutoff-)\bS_{-,p})\be_{0}\be_{1,-}^\Trans (f(\cutoff+)\bS_{+,p}+f(\cutoff-)\bS_{-,p})\\
&  \qquad- h  f(\cutoff+) F(\cutoff) (f(\cutoff+)\bS_{+,p}+f(\cutoff-)\bS_{-,p})\be_{1,+}\be_{0}^\Trans(f(\cutoff+)\bS_{+,p}+f(\cutoff-)\bS_{-,p})\\
&  \qquad+h \frac{F^{(2)}(\cutoff)}{f(\cutoff+)}(1-F(\cutoff))F(\cutoff) (f(\cutoff+)\bS_{+,p}+f(\cutoff-)\bS_{-,p})\be_{1,+}\be_{0}^\Trans (f(\cutoff+)\bS_{+,p}+f(\cutoff-)\bS_{-,p})\\
&  \qquad- h  f(\cutoff+) F(\cutoff) (f(\cutoff+)\bS_{+,p}+f(\cutoff-)\bS_{-,p})\be_{0}\be_{1,+}^\Trans (f(\cutoff+)\bS_{+,p}+f(\cutoff-)\bS_{-,p})\\
&  \qquad+h \frac{F^{(2)}(\cutoff)}{f(\cutoff+)}F(\cutoff)(1-F(\cutoff))  (f(\cutoff+)\bS_{+,p}+f(\cutoff-)\bS_{-,p})\be_{0}\be_{1,+}^\Trans (f(\cutoff+)\bS_{+,p}+f(\cutoff-)\bS_{-,p})\\
&  \qquad+hf(\cutoff-)  (f(\cutoff+)\bS_{+,p}+f(\cutoff-)\bS_{-,p})\be_{1,-}\be_{0}^\Trans (f(\cutoff+)\bS_{+,p}+f(\cutoff-)\bS_{-,p})\\
&  \qquad+h  f(\cutoff-)  (f(\cutoff+)\bS_{+,p}+f(\cutoff-)\bS_{-,p})\be_{0}\be_{1,-}^\Trans (f(\cutoff+)\bS_{+,p}+f(\cutoff-)\bS_{-,p})\\
&  \qquad+h  (f(\cutoff+)^3 \bGamma_{+,p}+  f(\cutoff-)^3 \boldsymbol{\Psi}\bGamma_{+,p}\boldsymbol{\Psi}).
\end{align*}
Therefore, 
\begin{align*}
&\ \Var\left[ (\be_{1,+}-\be_{1,-})'\sqrt{\frac{n}{h}}(f(\cutoff+)\bS_{+,p}+f(\cutoff-)\bS_{-,p})^{-1}\hat{\bL} \right]\\
&= (\be_{1,+}-\be_{1,-})'(f(\cutoff+)\bS_{+,p}+f(\cutoff-)\bS_{-,p})^{-1} 
(f(\cutoff+)^3 \bGamma_{+,p}\\
&\qquad+  f(\cutoff-)^3 \boldsymbol{\Psi}\bGamma_{+,p}\boldsymbol{\Psi})
(f(\cutoff+)\bS_{+,p}+f(\cutoff-)\bS_{-,p})^{-1}(\be_{1,+}-\be_{1,-}) + O(h).
\end{align*}
\end{proof}

\subsection{Proof of Lemma \ref{lem14: quadratic variance, restricted}}

\begin{proof}
This follows from Lemma \ref{lem4: quadratic variance} by splitting the bias calculation for the two subsamples, below and above the cutoff $\cutoff$. 
\end{proof}

\subsection{Proof of Corollary \ref{coro2: asy normal, restricted}}

\begin{proof}
This follows from the previous lemmas and verifying the Lindeberg condition. See also the proof of Lemma \ref{lem3: linear variance}, Theorem \ref{thm1: asy normality} and Theorem \ref{thm2: variance estimation}.
\end{proof}

}

\clearpage
\addcontentsline{toc}{section}{Tables}
\newcounter{K} % kernel
\newcounter{M} % evaluation point
\newcounter{P} % polynomial order

\newcommand\Kvval{1}
\newcommand\Mvval{1}
\newcommand\Pvval{1}

\forloop{K}{1}{\value{K} < 2} {
\forloop{M}{1}{\value{M} < 5} {
\forloop{P}{1}{\value{P} < 3} {

\ifthenelse{\theK=1}{\renewcommand\Kvval{triangular}}{\ifthenelse{\theK=2}{\renewcommand\Kvval{epanechnikov}}{\renewcommand\Kvval{uniform}}}

\ifthenelse{\theM=1}{\renewcommand\Mvval{-0.8}}{\ifthenelse{\theM=2}{\renewcommand\Mvval{-0.5}}{\ifthenelse{\theM=3}{\renewcommand\Mvval{0.5}}{\ifthenelse{\theM=4}{\renewcommand\Mvval{1.5}}{\renewcommand\Mvval{2}}}}}

\ifthenelse{\theP=1}{\renewcommand\Pvval{2}}{\renewcommand\Pvval{3}}

\clearpage
%\begin{landscape}
\begin{table}[tp]
\centering

\caption{Simulation (truncated Normal). $x=\Mvval$, $p=\Pvval$, \Kvval\ kernel. \bigskip}

\renewcommand{\arraystretch}{1.1}

\subfloat[$n=1000$]{\resizebox{0.48\textwidth}{!}{\scriptsize\input{simulation/Supplement/Normal/summary_\Mvval_\Pvval_\Kvval_1000_.txt}}} 
\hskip 0.03\textwidth
\subfloat[$n=2000$]{\resizebox{0.48\textwidth}{!}{\scriptsize\input{simulation/Supplement/Normal/summary_\Mvval_\Pvval_\Kvval_2000_.txt}}} \\ \medskip
{\resizebox{0.48\textwidth}{!}{\scriptsize\input{simulation/Supplement/Normal/summary_h_\Mvval_\Pvval_\Kvval_1000_.txt}}}
\hskip 0.03\textwidth
{\resizebox{0.48\textwidth}{!}{\scriptsize\input{simulation/Supplement/Normal/summary_h_\Mvval_\Pvval_\Kvval_2000_.txt}}}

{\scriptsize
\begin{flushleft}
\textbf{Note.} \underline{(i) bias}: empirical bias of the estimators; \underline{(ii) sd}: empirical standard deviation of the estimators; \underline{(iii) $\sqrt{\text{mse}}$}: empirical MSE of the estimators; \underline{(iv) mean}: empirical average of the estimated standard errors; \underline{(v) size}: empirical size of testing the hypothesis at nominal 5\% level, the test statistic is centered at $\Expectation \hat{f}_p$. 
\end{flushleft}
}

\end{table}
%\end{landscape}
\clearpage

}}}

\forloop{K}{1}{\value{K} < 2} {
\forloop{M}{1}{\value{M} < 4} {
\forloop{P}{1}{\value{P} < 3} {

\ifthenelse{\theK=1}{\renewcommand\Kvval{triangular}}{\ifthenelse{\theK=2}{\renewcommand\Kvval{epanechnikov}}{\renewcommand\Kvval{uniform}}}

\ifthenelse{\theM=1}{\renewcommand\Mvval{0}}{\ifthenelse{\theM=2}{\renewcommand\Mvval{1}}{\ifthenelse{\theM=3}{\renewcommand\Mvval{1.5}}{\ifthenelse{\theM=4}{\renewcommand\Mvval{1.5}}{\renewcommand\Mvval{2}}}}}

\ifthenelse{\theP=1}{\renewcommand\Pvval{2}}{\renewcommand\Pvval{3}}

\clearpage
%\begin{landscape}
\begin{table}
\centering

\caption{Simulation (Exponential). $x=\Mvval$, $p=\Pvval$, \Kvval\ kernel. \bigskip}

\renewcommand{\arraystretch}{1.1}

\subfloat[$n=1000$]{\resizebox{0.48\textwidth}{!}{\scriptsize\input{simulation/Supplement/Exponential/summary_\Mvval_\Pvval_\Kvval_1000_.txt}}} 
\hskip 0.03\textwidth
\subfloat[$n=2000$]{\resizebox{0.48\textwidth}{!}{\scriptsize\input{simulation/Supplement/Exponential/summary_\Mvval_\Pvval_\Kvval_2000_.txt}}} \\ \medskip
{\resizebox{0.48\textwidth}{!}{\scriptsize\input{simulation/Supplement/Exponential/summary_h_\Mvval_\Pvval_\Kvval_1000_.txt}}}
\hskip 0.03\textwidth
{\resizebox{0.48\textwidth}{!}{\scriptsize\input{simulation/Supplement/Exponential/summary_h_\Mvval_\Pvval_\Kvval_2000_.txt}}}

{\scriptsize
\begin{flushleft}
\textbf{Note.} \underline{(i) bias}: empirical bias of the estimators; \underline{(ii) sd}: empirical standard deviation of the estimators; \underline{(iii) $\sqrt{\text{mse}}$}: empirical MSE of the estimators; \underline{(iv) mean}: empirical average of the estimated standard errors; \underline{(v) size}: empirical size of testing the hypothesis at nominal 5\% level, the test statistic is centered at $\Expectation \hat{f}_p$. 
\end{flushleft}
}

\end{table}
%\end{landscape}
\clearpage

}}}